\documentclass[pr, twocolumn, preprintnumbers, showpacs, footnoteadded, superscriptaddress,longbibliography]{revtex4-1}
\pdfoutput=1

\usepackage[T1]{fontenc}
\usepackage{amssymb}
\usepackage{graphicx}
\usepackage{subfigure}
\usepackage{amsmath}
\usepackage{slashed}
\usepackage{tensor}
\usepackage{hyperref}
\usepackage{epstopdf}
\usepackage{extarrows}
\usepackage{xcolor}
\usepackage{enumerate}

\newcommand{\be}{\begin{equation}}
\newcommand{\ee}{\end{equation}}
\newcommand{\bea}{\setlength\arraycolsep{2pt} \begin{eqnarray}}
\newcommand{\eea}{\end{eqnarray}}

\newcommand{\bpm}{\begin{pmatrix}}
\newcommand{\epm}{\end{pmatrix}}

\def\fft#1#2{{\frac{#1}{#2}}}

\def\0{{\sst{(0)}}}
\def\1{{\sst{(1)}}}
\def\2{{\sst{(2)}}}
\def\3{{\sst{(3)}}}
\def\4{{\sst{(4)}}}
\def\5{{\sst{(5)}}}
\def\6{{\sst{(6)}}}
\def\7{{\sst{(7)}}}
\def\8{{\sst{(8)}}}
\def\sst#1{{\scriptscriptstyle #1}}

\def\der{{ \mathrm{d} }}
\def\hrn{{ h_{\text{RN}} }}

\def\mrn{{ m_{\text{RN}} }}

\def\rhobar{ { \bar{\rho} }}

\usepackage{changes}
\usepackage{braket}
\definechangesauthor[name={Per cusse}, color=orange]{per}

\begin{document}

\title{\boldmath  Asymptotically-flat Black holes in Bumblebee gravity:\\ 
Exact solutions and Thermodynamics}
\author{Jinbo Yang}
\email{yangjinbo@gzhu.edu.cn}
\affiliation{Department of Astronomy, School of Physics and Materials Science,\\
				Guangzhou University, Guangzhou 510006, China}
\author{Zhan-Feng Mai}
\email{zhanfeng.mai@gmail.com}
\affiliation{Guangxi Key Laboratory for Relativistic Astrophysics, School of Physical Science and Technology, Guangxi University, Nanning 530004, China}
\author{Dicong Liang}
\affiliation{Department of Mathematics and Physics, School of Biomedical Engineering, Southern Medical University, Guangzhou, 510515, China}
\author{Lijing Shao}
\affiliation{Kavli Institute for Astronomy and Astrophysics, Peking University, Beijing 100871, China}%
\affiliation{National Astronomical Observatories, Chinese Academy of Sciences, Beijing 100101, China}

\begin{abstract}
We construct analytic solutions to the bumblebee gravity theory in static and spherically symmetric spacetimes, where the bumblebee vector field admits only a non-vanishing temporal component. 
In particular, we identify the parameter space that allows for asymptotically flat black hole solutions. We further investigate the thermodynamic properties of these black holes and obtained the analytic formulas for the $Y$ charge and $X$ potential, which were introduced in the prior work to ensure the Smarr relation and the first law of black hole thermodynamics. 
Using the new analytic results, we verify the numerical findings reported in early work and uncover multiple cases missed in the previous numerical analysis. 
These include: (i) an unbounded charge-mass ratio when the non-minimal coupling parameter \(\xi\) is larger than \(2\kappa\), (ii) the emergence of a traversable wormhole configuration for overcharged solutions with \(\xi<0\), (iii) the non-monotonic turning behavior of the Hawking temperature as a function of the charge-mass ratio, and (iv) the presence of two divergent points in the constant-$Y$ heat capacity..
\end{abstract}

\maketitle
\flushbottom

\allowdisplaybreaks

\section{Introduction}
Spontaneous Lorentz symmetry breaking is a theoretical hypothesis wherein a tensor-valued operator acquires a non-zero vacuum expectation value (VEV), thus defining a preferred spacetime direction. This phenomenon can be realized in certain quantum gravity theories \cite{Kostelecky:1988zi, Kostelecky:1989jp, Kostelecky:1989jw, Kostelecky:2003fs}. 
To capture the low-energy predictions arising from such a tensor-valued VEV, an effective field theory (EFT) framework known as the Standard Model Extension (SME) has been proposed \cite{Colladay:1996iz, Kostelecky:2003fs, Lehnert:2023wzr}.
As an EFT, all low-energy operators constructed from Standard Model fields, along with the tensor fields that acquire VEVs, must be systematically considered.
On the other hand, since the ultraviolet (UV) completion at high energy scales may correspond to a quantum gravity theory, it is natural to anticipate that the tensor fields with non-zero VEVs in the low-energy EFT will exhibit non-trivial couplings to gravitational degrees of freedom.
The bumblebee theory is the simplest vector-tensor model with such couplings that yields several non-trivial physical consequences~\cite{Kostelecky:1989jp, Kostelecky:2003fs, Xu:2022frb}. 
This theory is characterized by a vector field $B_{\mu}$ termed the ``bumblebee field'', whose non-zero VEV induces spontaneous Lorentz symmetry breaking. This vector field is non-minimally coupled to the Ricci tensor. 
In the weak-field approximation, the vacuum configuration, consisting of Minkowski spacetime and a constant non-zero VEV $B_{\mu}=b_{\mu}$ for the bumblebee field, has been shown to be stable in some parameter space \cite{Mai:2024lgk}.

From the perspective of modified gravity, the bumblebee theory can be categorized as a vector-tensor theory, enabling the investigation of strong-field effects through the solution of classical equations of motion (EoMs). \citet{Casana:2017jkc} pioneered the study of black-hole solutions in bumblebee gravity by deriving a Schwarzschild-like solution. Since then, a variety of vacuum spacetimes have been constructed in the bumblebee theory and its extensions \cite{Xu:2022frb,Filho:2022yrk,Ding:2023niy,Lessa:2023yvw,AraujoFilho:2024ykw,Liu:2024axg,Xu:2025jvk,Liu:2025oho,Chen:2025ypx,Bailey:2025oun,Li:2025rjv, Zhu:2025fiy,Zhu:2026vae}. 
Traversable wormhole solutions are admitted in bumblebee gravity and they do not require exotic matter violating the null energy condition \cite{Ovgun:2018xys}.
In addition, matter-supported solutions, such as neutron stars and quark stars have also been investigated in Refs.~\cite{Ji:2024aeg,Panotopoulos:2024jtn,Neves:2024ggn}.
The rich spectrum of compact-object solutions has motivated extensive investigations into their physical and observational properties.
In particular, the thermodynamics \cite{Mai:2023ggs,An:2024fzf} and quasi normal modes \cite{Liu:2022dcn,Liu:2024oeq,Li:2025itp} of bumblebee black holes have been studied in details.
Current constraints on the theory have been obtained from black-hole imaging observations \cite{Xu:2023xqh,Lambiase:2024uzy} and measurements of the S2-star orbit \cite{QiQi:2026pnb}.
While future observations, including pulsar timing measurements \cite{Hu:2023vsg} and gravitational wave detection \cite{Liang:2022hxd,Liang:2022gdk,Liu:2025swi,Lai:2025nyo,Shi:2026zxx}, are expected to provide complementary and more stringent tests. 
The implications of bumblebee gravity are not limited to compact objects, 
a variety of scenarios on cosmological scale are also studied \cite{Maluf:2021lwh,Khodadi:2022mzt,Sarmah:2024xwx,Zhu:2024qcm,Xu:2025heq}.

Previously, \citet{Xu:2022frb} have constructed asymptotically flat bumblebee black holes via a systematic numerical approach. 
The behavior of light propagation in these bumblebee spacetimes forms the basis for observational consequences with electromagnetic probes and has also been systematically investigated. 
Since then, Refs. \cite{ Xu:2023xqh, Hu:2023vsg} have utilized the M87* black hole image and other astrophysical observations to constrain the physical parameters of these bumblebee black holes. 
From a theoretical perspective, the thermodynamics of these numerically derived asymptotically flat solutions have also been explored by \citet{Mai:2023ggs}.
A notable finding is that the bumblebee charge associated with the $B_{\mu}$ field cannot be directly incorporated into the first law of thermodynamics or the Smarr relation. 
Instead, these two fundamental relations are satisfied by the newly constructed $Y$ charge and $X$ potential. Based on these results, the local thermodynamic stability has been analyzed by calculating the heat capacity for constant-$Y$ processes \cite{Mai:2023ggs}. Subsequent studies on numerically derived bumblebee black holes have focused on their dynamical stability \cite{Mai:2024lgk}. However, accumulated numerical errors remain a major obstacle to fully scanning the parameter space, and open questions such as the nature of the upper limit for the charge-mass ratio remain unresolved.
The goal of this study is to overcome the limitations of previous numerical studies by leveraging recently constructed exact analytic solutions, which explicitly include asymptotically flat black hole configurations.

The paper is organized as follows. 
Section~\ref{sec:theory} provides a brief overview of the action and EoMs of the bumblebee theory, focusing on static spherically symmetric spacetimes and the scenario where the bumblebee field has only a time-component. We then present a family of exact solutions that explicitly includes black hole cases.
In Section~\ref{sec:BHs} we establish a systematic method to verify asymptotic flatness and identify the parameter space corresponding to black hole solutions. We emphasize that solutions with identical analytic expressions may encompass parameter regimes corresponding to traversable wormholes rather than black holes. Thus, a primary task is to systematically isolate the parameter regime that yields black holes.
Section~\ref{sec:thermo} discusses black hole thermodynamics. Using the analytic expressions, we confirm results from previous literature and address gaps in the earlier investigations. Finally, our conclusions are given in Section~\ref{sec:con}.

\section{The Bumblebee theory and its exact solutions} \label{sec:theory}
\subsection{Basic equations of the bumblebee theory}

The action of the bumblebee gravity model is given by~\cite{Kostelecky:2003fs}
\begin{widetext}
\begin{eqnarray}
		I
		&& =  \int \sqrt{-g}\der^4x \left( \frac{1}{2\kappa} R +\frac{\xi}{2\kappa} R^{\mu\nu} B_{\mu}B_{\nu} 
        - \frac{1}{4}B_{\mu\nu}B^{\mu\nu} -V( B_{\mu}B^{\mu} \pm b^2) \right)  \,,
\end{eqnarray}
where $ B_{\mu\nu}	=  \nabla_{\mu}B_{\nu} - \nabla_{\nu}B_{\mu} $ is the strength tensor of the bumblebee vector field $B_{\mu}$, $\nabla_{\mu}$ denotes the covariant derivative satisfying $\nabla_{\mu}g_{\nu\lambda}=0$, and $g$ is the determinant of $g_{\mu\nu}$.
Then, the corresponding EoMs are
\begin{align}
	&R_{\mu\nu} - \frac{1}{2}Rg_{\mu\nu} =\kappa T^B_{\;\mu\nu}  \label{eomgra}\,,\\
	& \nabla_{\nu} B^{\mu\nu}=\frac{\xi}{\kappa} R^{\mu\nu}B_{\nu} -2V'B^{\mu} \label{eomvec}\,,
\end{align}
where the effective energy-momentum tensor $T^{B}_{\mu\nu}$ from the bumblebee field is
\begin{align}
	T^{B}_{\mu\nu} =& ~ B_{\mu \lambda}B_\nu{}^{\lambda}-g_{\mu\nu}\left(\frac{1}{4}B^{\alpha\beta}B_{\alpha\beta} +V\right) + 2V'B_{\mu}B_{\nu} 
    - \frac{\xi}{2\kappa} \bigg( \nabla_{\lambda}\nabla^{\lambda} ( B_\mu B_\nu ) + g_{\mu\nu} \nabla_\alpha \nabla_\beta ( B^\alpha B^\beta ) \nonumber \\
   & - \nabla_\lambda \nabla_\mu ( B^\lambda B_\nu ) - \nabla_\lambda \nabla_\nu ( B_\mu B^\lambda )  
   + 2 B_\mu B_\lambda R_\nu^{\; \lambda}  +  2 B_\nu B_\lambda R_\mu^{\; \lambda}  -g_{\mu\nu} B^\alpha B^\beta R_{\alpha\beta}  \bigg) \,,
\end{align}
and $V'={\der V}/{\der(B_\mu B^\mu)}$.

We suppose that specific bumblebee field configurations $B_{\mu}=b_{\mu}$ minimize the potential, i.e. $V=V'=0$, and allow for non-trivial spacetime geometries that are exact solutions to Eqs. \eqref{eomgra} and \eqref{eomvec}.
More concretely, we consider the following ansatz:
\begin{align}
	&\der s^2=-h(\rho)\der t^2+\fft{\der \rho^2 }{h(\rho)}  +r^2(\rho)\der\Omega^2\,,\nonumber \\
	& B_{\mu}\der x^{\mu} = b_t(\rho) \der t \,, \label{ansatz}
\end{align}
where $\der\Omega^2=\der\theta^2 + \sin^2\theta \der\phi$.
Then, Eq.~\eqref{eomgra} yields the following ordinary differential equations (ODEs):
\begin{align}
	0 = E_{tt} =&  3\xi b_t^2 r^4 {h'}^2 - 4r^2 h^3 \big( {r'}^2 + 2r r'' \big) +2h^2 r^2 \big(2+4r\xi b_t b_t' r' - 2r r' h' + r^2\big( 2\xi b_t b_t'' - (\kappa-2\xi) {b_t'}^2 \big)\big)  \nonumber \\ 
    & -2\xi b_t r^2 h \big( 4 r b_t h' r' + r^2 \big( 3b_t' h' +2b_t h''\big)\big)  \,, \label{odegratt}\\
	0 = E_{\rho\rho} =& 2\xi b_t h r^4 b_t' h' -\xi b_t^2 r^4 {h'}^2 - 4r^2 {r'}^2 h^3  
    -2h^2 r^2 \big(\kappa r^2 {b_t'}^2 +2r r' h'-2 \big)  \label{odegraxx} \,,\\
	0 = E_{\theta\theta} =& 2\xi b_t h r^4 b_t' h' -\xi b_t^2 r^4 {h'}^2  - 4 r^2 {r'}^2 h^3 
    +2h^2 r^2 \big( -\kappa r^2 {b_t'}^2 +2r r' h' + r^2 h'' + 2h\big({r'}^2 + r r''\big)\big) 
    \label{odegraang} \,,
\end{align}
where the prime denotes $\der/\der \rho$, and the $\phi$-$\phi$ component yields the same equation as the $\theta$-$\theta$ component, Eq. \eqref{odegraang}. Meanwhile, Eq. \eqref{eomvec} gives
\be
h \fft{\der \left( r^2\,b_t' \right)}{\der \rho}  = \fft{\xi}{2\kappa} \, b_t \fft{\der\left(r^2\, h'\right)}{\der \rho} . \label{fromeomvec}
\ee

Appendix \ref{appendix} provides an introduction to constructing general exact solutions for Eqs.~\eqref{odegratt}, \eqref{odegraxx}, \eqref{odegraang}, and \eqref{fromeomvec}.
The main strategy is to construct the solution for the case when $-b_{t}^2/h$ is a constant,  which was also obtained in Ref.~\cite{Li:2025rjv, Zhu:2025fiy}. 
Consistency requires that the constant should be $2/\kappa$. Such a feature suggests an alternative choice of radius coordinate, which can be regarded as the misalignment between $-b_{t}^2/h$ and $2/\kappa$. Then, adjusting the ansatz to fit the new coordinate yields an integrable family of solutions. 
A further choice of specific integral constants includes exact solutions in terms of elementary functions.

\subsection{Exact solutions including black holes}

Then we focus on the exact solutions, which may include black hole situations:
\begin{align}	
h(\rho)=&~\beta^2 s(s-2)\frac{u(\rho)-1}{u(\rho)+1} \big(u(\rho) + s-1 \big)^{2(s-1)} \,, \label{BHlapsefunconfig} 
	\\ 	
    r^2(\rho) =& \frac{\rho^2-a^2}{h(\rho)} 
    = \frac{\rho^2-a^2}{\beta^2s(s-2)}\frac{u(\rho)+1}{u(\rho)-1} \big(u(\rho) + s-1 \big)^{2(1-s)} \label{BHrsqconfig}  \,,	
	\\	b_t(\rho)=& \beta \sqrt{\frac{2}{\kappa}} \big(u(\rho)-1\big)\big( u(\rho) + s-1\big)^{s-1}  \label{BHb0config} \,,
\end{align}
with function $u(\rho)$  defined through
\begin{align}
	\rho(u)= a \frac{ u+1 + P e^{2u} (u-1) }{ u+1 - P e^{2u} (u-1) }  \,,  \label{BHfamxofu}
\end{align}
where we introduce $s=\xi/\kappa$ for short; $a$ is an integral constant with length dimension; $\beta$ and $P$ are dimensionless integral constants. 
For clearer expression using elementary functions, these solutions can alternatively be expressed in the  $\{t,u,\theta,\phi\}$ coordinate system instead of $\{t,\rho,\theta,\phi\}$:
\begin{align}
	\der s^2 =& - \bigg( \beta^2 s(s-2)\frac{u-1}{u+1} \big(u + s-1 \big)^{2(s-1)} \bigg) \der t^2 
    +\frac{16 a^2 P^2}{\beta^2s(s-2)} \frac{ e^{4u} u^4 (u+s-1)^{-2(s-1)} (u+1)}{\big( u+1-Pe^{2u}(u-1) \big)^4 (u-1)}  \der u^2
	\nonumber \\ 
	& + \frac{4 a^2 P}{\beta^2s(s-2)} \frac{ e^{2u}(u+s-1)^{-2(s-1)} (u+1)^2 }{\big( u+1-Pe^{2u}(u-1) \big)^2} \der\Omega^2  
	\,,\nonumber\\
	b_t(u) \der t =& \beta \sqrt{\frac{2}{\kappa}} \big(u-1\big)\big( u + s-1\big)^{s-1}  \der t \,.\label{BHfam}	
\end{align}
\end{widetext}
A correct metric signature requires $Ps(s-2)>0$.
Moreover, it is worth noting that the solution \eqref{BHfam} contains only two independent parameters. This is because the solution retains the same form under the following rescaling,
\begin{align}
	\beta = l\tilde{\beta} \,,\quad
	a = l \tilde{a} \,,\quad
	t =  \tilde{t}/l \,,\quad 
	\rho = l \tilde{\rho}  \,.
\end{align}
Hence, one can always adjust the value of $\beta$ for the same solution.

The above solutions \eqref{BHfam} include black hole configurations, justifying their classification as a charged black hole solution family. In particular, the presence of a Killing horizon at $u=1$ guarantees this property. Meanwhile, the areal radius $r_H$ and the surface gravity $\kappa_H=h'/2$ for the Killing horizon $u=1$ are
\begin{align}
	&~r_H = 2a e^2|s|^{-s}  \sqrt{\frac{s}{s-2} \frac{P}{\beta^2} } \,,\\
	&~\kappa_H = \frac{1}{2}\bigg[\frac{\der h/\der u}{\der \rho/\der u}\bigg]_{u=1} = \frac{|s|^{2s}}{4a e^2} \frac{s-2}{s} \frac{\beta^2 }{P}\,,
\end{align}
which satisfy a simple relation
\begin{align}
	\kappa_H \, r_H^2  = a \,.\label{kappaXrHsq}
\end{align}
It would be worth noting that a Killing horizon is not necessarily a black hole horizon, but would probably be a hypersurface resembling the cosmic horizon of de Sitter spacetime.  In addition, the asymptotic flatness inherent in Eq.~\eqref{BHfam} is not directly accessible, 
complicating the application of the traditional Arnowitt-Deser-Misner (ADM) formalism for calculating mass and charge parameters. 
We will develop a specialized computational approach to address these limitations in the next section.

\section{Asymptotically flat black holes} \label{sec:BHs}
In this section, we will develop a systematic approach to confirm the asymptotically flatness of the solution \eqref{BHfamxofu}, and identify the parameter regime that corresponds to the charged black hole. 
\subsection{Asymptotical flatness}

\begin{figure}[t!]
	\begin{center}
		\includegraphics[width=0.235\textwidth]{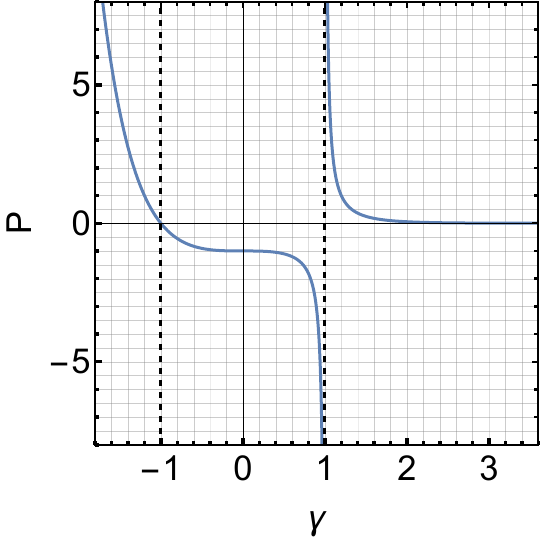} 
	\end{center}
	\caption{\small Relation between $\gamma$ and $P$.  If $P>0$, there exist two solutions for $\gamma$: one greater than 1 and the other smaller than $-1$; if $P<0$, the range of $\gamma$ is $-1<\gamma<1$.
    }
	\label{PVSgammaplt}  
\end{figure}

\begin{figure}[t!]
	\begin{center}
		\includegraphics[width=0.23\textwidth]{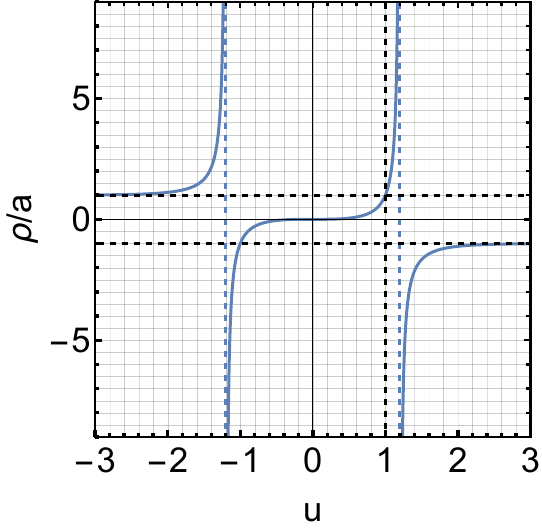}
		\includegraphics[width=0.235\textwidth]{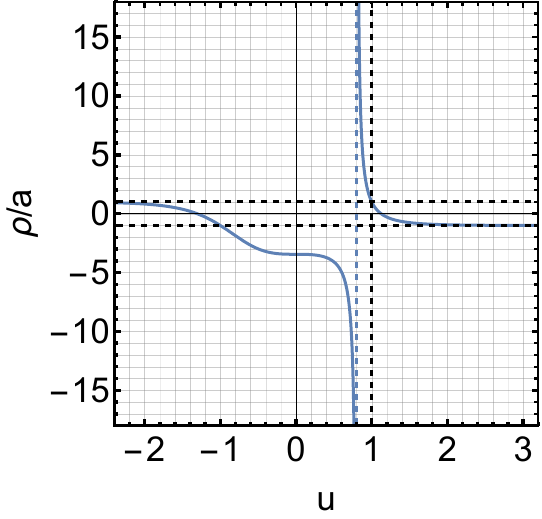} 
	\end{center}
	\caption{Examples of $\rho(u)$. Left: $\gamma=1.2$, $P= 0.9979$, the alternative value for $\gamma$ is $\gamma_{\text{alt}}=-1.1994$ ;
		Right: $ \gamma=0.8$, $P=-1.8170$. } 
	\label{rhoperaVSu}  
\end{figure}

As a preliminary step, we consider the line element in the coordinate system $\{t,r,\theta,\phi\}$, which reads:
\begin{align}
	&\der s^2=-h(r)\der t^2+\fft{\der r^2 }{f(r)}  +r^2\der\Omega^2\,. 
\end{align}
Asymptotic flatness is guaranteed under the following conditions:
\begin{align}
	\lim_{r\rightarrow \infty}h(r)=\lim_{r\rightarrow \infty}f(r) =1 \label{asymflatConditions}\,. 
\end{align}
While the requirement on $f(r)$ is well-posed, the condition on $h(r)$ can be relaxed. It is sufficient to require that $\lim_{r\rightarrow\infty}h(r)$ is a positive finite value, as the time coordinate $t$ can always be rescaled to normalize this limit.
For consistency with asymptotic flatness, $\rho$ must also tend to infinity as $r\rightarrow\infty$ to ensure $r\sim \rho$. 
Notably, both Eqs.~\eqref{BHfamxofu} and \eqref{BHfam} contain the factor $u+1-Pe^{2u}(u-1)$ in their denominators. To identify the asymptotic boundary of the spacetime, we introduce a parameter $\gamma$ satisfying
\begin{align}  
	\gamma +1 - P e^{2\gamma} (\gamma-1) =0
	\,, \label{rhopoleeq}
\end{align}
such that both $r(u)$ and $\rho(u)$ diverge at $u=\gamma$. Furthermore, Eq.~\eqref{rhopoleeq} enables us to express the integral constant $P$ as
\begin{align}
	P = \frac{\gamma+1}{\gamma-1}e^{-2\gamma} 
	\label{Pofgamma}\,.
\end{align}
The function $P(\gamma)$, i.e. Eq.~\eqref{Pofgamma}, is plotted in Fig. \ref{PVSgammaplt}. For positive $P$, two distinct solutions for $\gamma$ exist, dividing the domain of $\rho(u)$ into three connected components. In contrast, a negative $P$ corresponds to a unique $\gamma$ within the interval $-1<\gamma<1$, such that $u<\gamma$ and $u>\gamma$ form two separate connected domains.
We further calculate $\der\rho/\der u$, which is given by
\begin{align}
	\frac{\der\rho}{\der u} = P \bigg( \frac{2 u e^u}{u+1 - P e^{2u}(u-1) } \bigg)^2  \,.
\end{align}

Consequently, within each connected domain, $\rho(u)$ increases monotonically for $P>0$ and decreases monotonically for $P<0$.
Examples of $\rho(u)$ for $\gamma=1.2$ and $\gamma=0.8$ are presented in Fig.~\ref{rhoperaVSu}. The left panel shows that $\rho$ diverges at $u=1.2$ (the specified $\gamma$-value) and $u=-1.1994$ (the other pole corresponding to the same $P$). Within the three connected components defined by these poles, $\rho(u)$ increases monotonically over the intervals $u<-1.1994$, $-1.1994<u<1.2$, and $u>1.2$. The right panel demonstrates that $\rho(u)$ decreases monotonically for both $u<0.8$ and $u>0.8$, with divergence occurring at $u=0.8$.

Suppose an appropriate value of $\gamma$ is chosen. To ensure $h=1$ at $u=\gamma$, the parameter $\beta^2$ can be rescaled as
\begin{align}	
	\beta^2 =&~ \frac{1}{s(s-2)} \frac{\gamma+1}{\gamma-1} \big(\gamma + s-1 \big)^{2(1-s)} \,, \label{beta2normalizedtheh}
\end{align}
which should be positive, such that one should take $\gamma>1$ for cases of $s<0$ and $s>2$, but take
$\gamma<1$ for $0<s<2$. These are consistent with the condition $Ps(s-2)>0$ to ensure a correct metric signature.
Furthermore, replacing $\beta^2$ by Eq. \eqref{beta2normalizedtheh}, $h(u)$ becomes 
\begin{align}	
	h =&~ \frac{u-1}{\gamma-1} \frac{\gamma+1}{u+1} \bigg( \frac{u+ s-1}{\gamma +s-1} \bigg)^{2(s-1)} \,. \label{BHlapsefunAdjusted}
\end{align}
Hence, we obtain an expression of $h$ which is normalized as $\rho$ tends to positive infinity.

It is still necessary to study $r^2/\rho^2$ and $f$ as $\rho\rightarrow +\infty$ to confirm the asymptotical flatness. As functions of $u$, they can be written as 
\begin{widetext}
\begin{align}
	\frac{r^2}{\rho^2} =&~ \bigg(\frac{ 2 (\gamma-1) (u+1) }{ e^{\gamma-u} (\gamma-1) (u+1) + e^{u-\gamma} (\gamma+1)(u-1) } \bigg)^2 \bigg(\frac{u+ s-1}{\gamma +s-1}\bigg)^{2(1-s)} 
	\,,\\
	f =&~ h \bigg( \frac{\der r}{\der\rho} \bigg)^2
	= \frac{h}{4r^2} \bigg( \frac{\der r^2}{\der u}\bigg)^2 \bigg( \frac{\der \rho}{\der u} \bigg)^{-2}  = \frac{u-1}{u+1} 
     \frac{ \big[ e^{-u+\gamma} (u+1)^2(\gamma-1) + e^{u-\gamma}(u^2 +2(s-1)u +1) \big]^2 }{ 4(\gamma^2-1) u^2 (u+s-1)^2 }
	\,,
\end{align}
\end{widetext}
in which the scaling~\eqref{beta2normalizedtheh} is applied. 
Their Taylor expansions up to the first order in the vicinity of $u=\gamma$ are
\begin{align}
	&\frac{r^2(u)}{\rho^2(u)} \simeq  1 - \frac{2\gamma(u-\gamma ) [(s-1)\gamma+1] }{(\gamma+s-1)(\gamma+1)(\gamma-1)} +\cdots  \,,\\
	&f \simeq  1 + \frac{2\gamma}{\gamma^2-1}\frac{  (s-1)\gamma+1}{\gamma+s-1} (u-\gamma) +\cdots   \,.
\end{align}
Both of $r^2/\rho^2$ and $f$ become $1$ as $u\rightarrow\gamma$, i.e., $\rho\rightarrow +\infty$. 
These expansions verified the limit \eqref{asymflatConditions}, which confirmed the asymptotical flatness of the solution \eqref{BHfam}.

\begin{figure}[t]
	\centering
	\subfigure[$\xi<0$\label{raQbyMpltximinus}]{
		\includegraphics[width=0.2\textwidth]{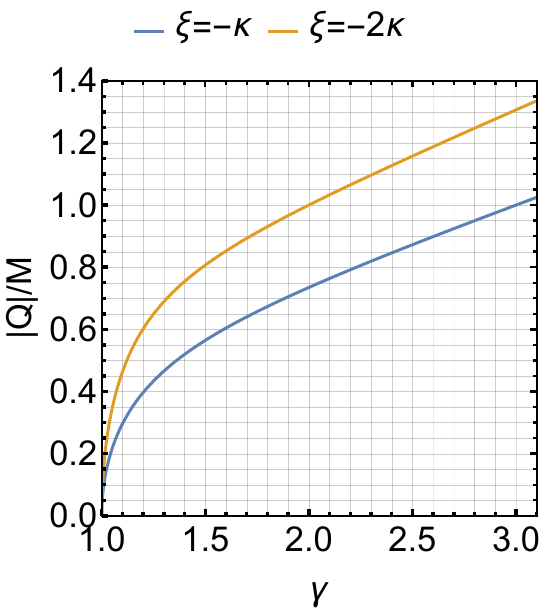}
	}		 
	\subfigure[$0<\xi<\kappa$ \label{raQbyMpltxi0to1} ]{
		\includegraphics[width=0.2\textwidth]{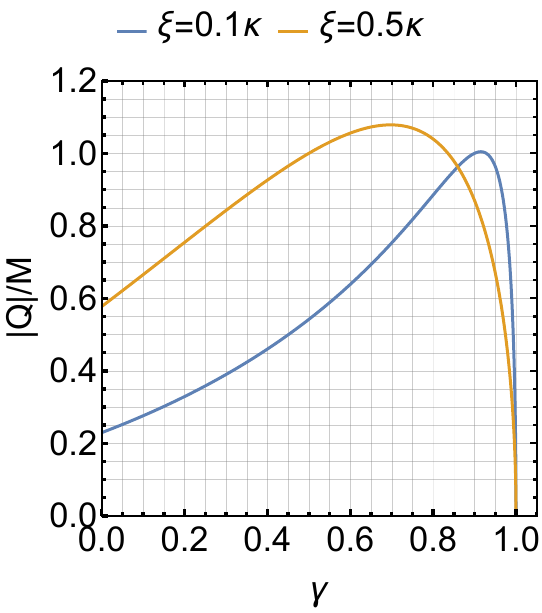} 
	}
	\subfigure[$\xi=\kappa$ \label{raQbyMpltxi1} ]{
		\includegraphics[width=0.205\textwidth]{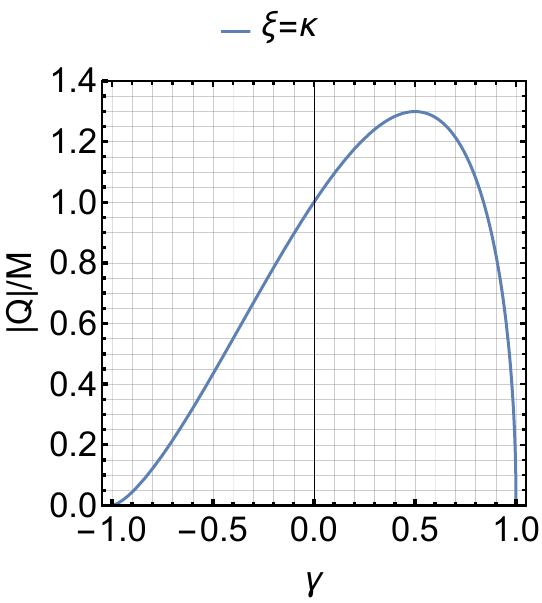}
	}
	\subfigure[$\kappa<\xi<2\kappa$ \label{raQbyMpltxi1to2} ]{
		\includegraphics[width=0.19\textwidth]{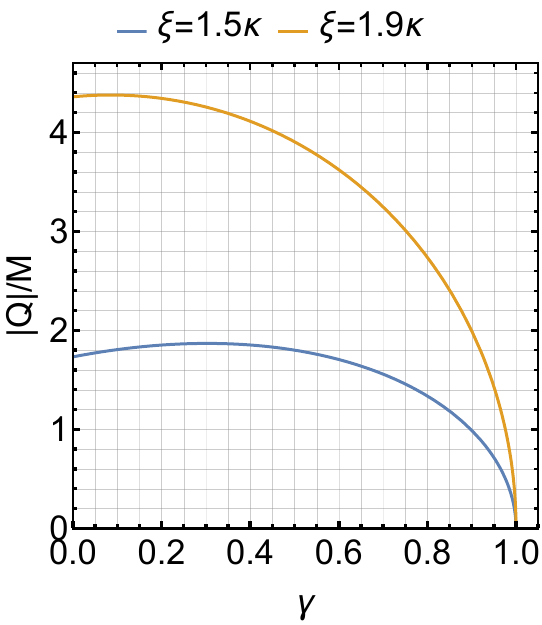}
	}
	\subfigure[$\xi>2\kappa$ \label{raQbyMpltxilarger2} ]{
		\includegraphics[width=0.2\textwidth]{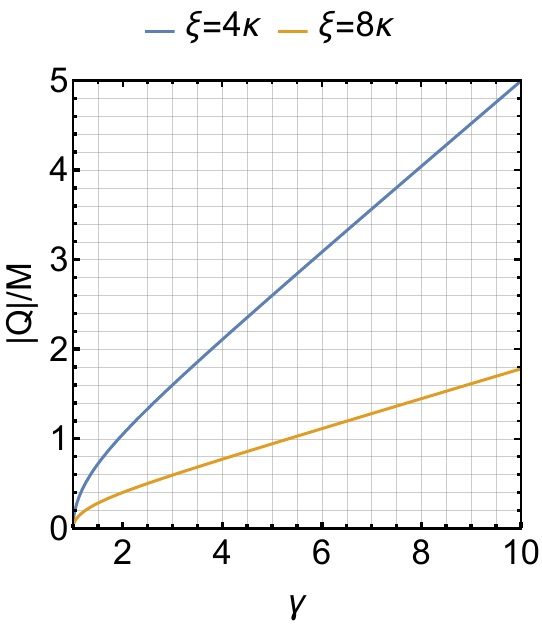}
	}	
\caption{\small Benchmarks for $|Q|/M$ versus $\gamma$ for various $\xi$.
}\label{gammaVSQplts}
\end{figure}

We then read the vector charge, ADM mass, and the asymptotic value of $b_t$ by Taylor expanding $b_t$ and $h$:
\begin{align}
	&M = \frac{a}{ \gamma } \frac{ (s-1)~ \gamma +1 }{ \gamma +s-1 } \,, \label{ADMmassbyagam}\quad \\
	&|Q| = \bigg|\frac{a}{ \gamma } \frac{\gamma+1}{\gamma+s-1} \bigg| \sqrt{ \frac{s (\gamma^2 -1) }{s-2}  }	\,, \quad \\
	&|b_{t,\infty}| = \sqrt{\frac{2}{\kappa }  \frac{\gamma^2-1}{s(s-2)}   }  \,.
\end{align}
We obtain the following ratios:
\begin{align}
	\frac{|Q|}{M} =&~ \bigg|\frac{\gamma+1}{(s-1)~ \gamma +1} \bigg| \sqrt{ \frac{s (\gamma^2 -1) }{s-2}  }	
	\,, \quad 
	\\
	\frac{r_H}{M} =&~ \frac{ 2 e^{1-\gamma  } }{(s-1)\gamma+1} |s|^{1-s} \gamma (\gamma-1+s)^s \,, 
\end{align}
which are controlled by the dimensionless parameter $\gamma$ when the value of $s$ is given. 
It would be beneficial to show the relation curve for the ratio $|Q|/M$ and $\gamma$ here (see Fig. \ref{gammaVSQplts}).
In the later subsection, we would discuss how the range of $\gamma$ affects the upper bound of the ratio $|Q|/M$.
Before exploring the physical parameter space, we still need to ensure whether there is an additional constraint on the range of $\gamma$ required for the solutions to describe black holes.

\subsection{Allowed range of $\gamma$ from the attractiveness of gravity}

\begin{figure}[hbpt]
	\centering
	\subfigure[$\xi<0$\label{hperbsqximinus}]{
		\includegraphics[width=0.22\textwidth]{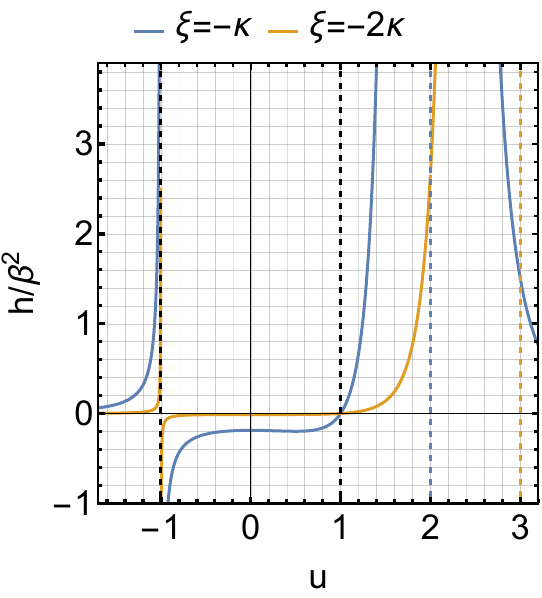}
	}		  
	\subfigure[$0<\xi<\kappa$ \label{hperbsqxi0to1} ]{
		\includegraphics[width=0.22\textwidth]{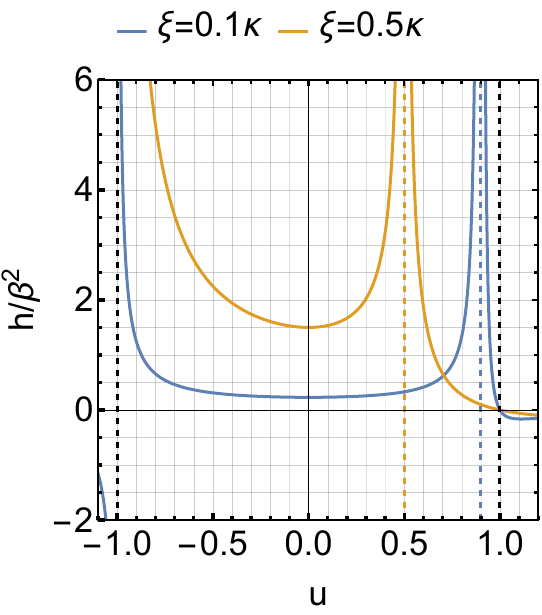} 
	}
	\subfigure[$\xi=\kappa$ \label{hperbsqxi1} ]{
		\includegraphics[width=0.225\textwidth]{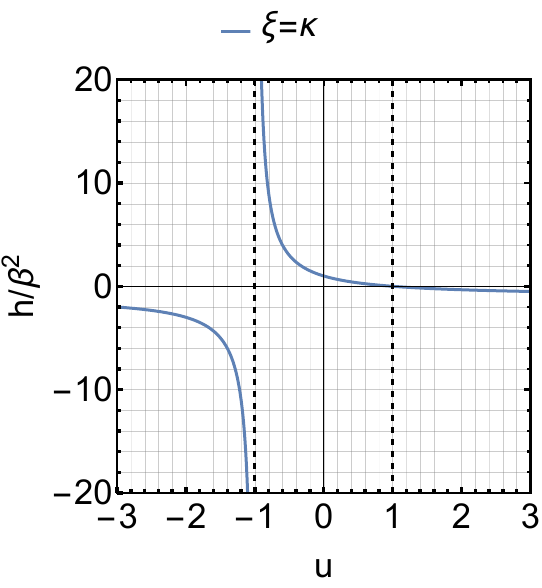}
	}
	\subfigure[$\kappa<\xi<2\kappa$ \label{hperbsqxi1to2} ]{
		\includegraphics[width=0.225\textwidth]{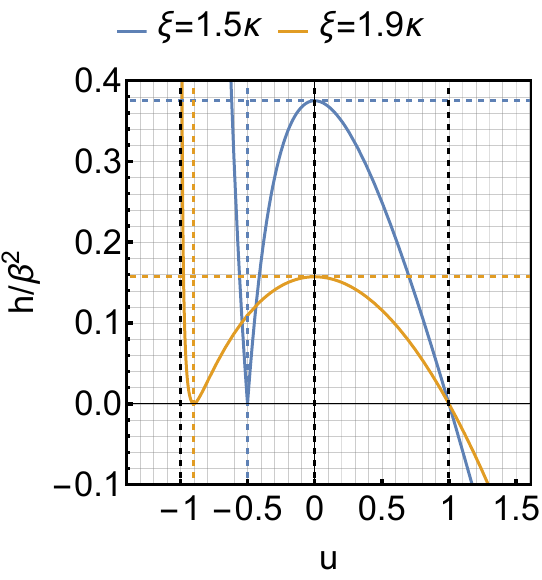}
	}
	\subfigure[$\xi=4\kappa$ \label{hperbsqxi4} ]{
		\includegraphics[width=0.22\textwidth]{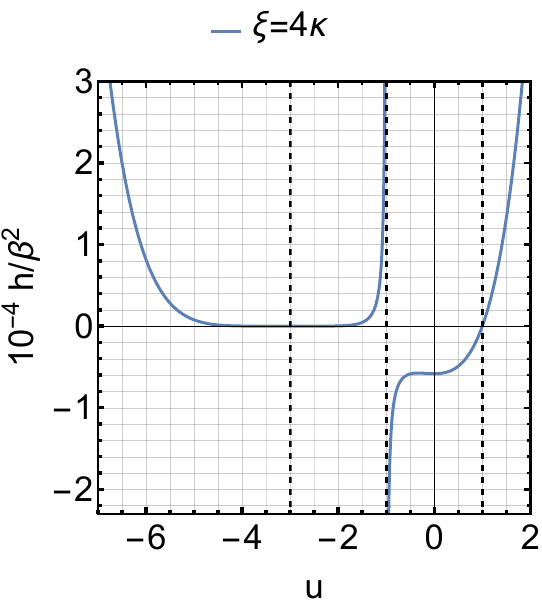}
	}
	\subfigure[$\xi=8\kappa$ \label{hperbsqxi8} ]{
		\includegraphics[width=0.23\textwidth]{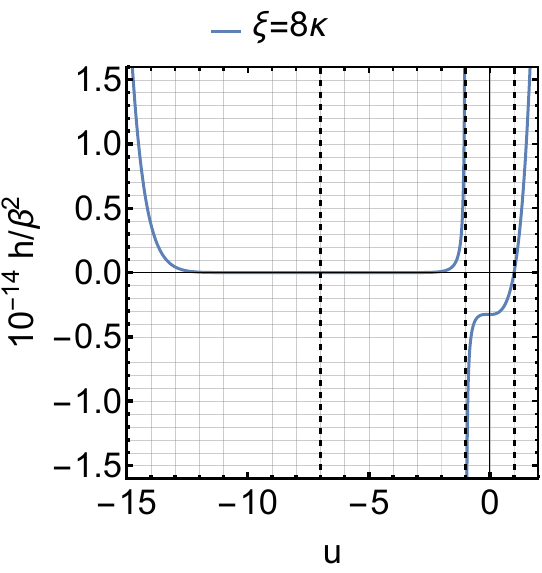}
	}
\caption{ Benchmarks for $h/\beta^2$ as a function of $u$ for various $\xi$. All of them contain the root $u=1$ and diverge at $u=-1$. Panel (a) shows typical cases with $\xi<0$, blue for $\xi=-\kappa$ and yellow for $\xi=-2\kappa$. The quantity $h/\beta^2$ diverges at $u=1-\xi/\kappa$, and increases monotonically in the region $-1<u<1-\xi/\kappa$ ; Panel (b) shows typical cases for $0<\xi<\kappa$, blue for $\xi=0.1\kappa$ and yellow for $\xi=0.5\kappa$. The quantity $h/\beta^2$ decreases monotonically in the region $u>1-\xi/\kappa$; Panel (c) shows typical cases for $\xi=\kappa$. The function $h/\beta^2$ contains the root $u=1$ and diverges at $u=-1$. It decreases monotonically in each connected domain. Panel (d) shows typical cases for $\kappa<\xi<2\kappa$, blue for $1.5\kappa$ and yellow for $1.9\kappa$. The function $h/\beta^2$ contains additional roots at $u=1-\xi/\kappa$. Another common feature of these cases is that $h/\beta^2$ generally exhibits a peak at $u=0$. We infer that the region $0<u<1$ corresponds to the exterior of the black hole.    Panels (e) and (f) show typical cases for $\xi>2\kappa$. While $u=1-\xi/\kappa$ is an additional root of $h/\beta^2$, it lies in a different connected domain from $u=1$. We infer that the region $u>1$ corresponds to the exterior of black holes. 
} 
\end{figure}

The behavior of $h(u)$ with different coupling $\xi$ is important to constrain the range of $\gamma$, such that the object described by solution \eqref{BHfam} contains a positive ADM mass. 
For convenience, to omit the less important overall factor of $h$, we consider the function $h/\beta^2$ given by
\begin{align}	
	\frac{h}{\beta^2}=&~ s(s-2)~\frac{u-1}{u+1} ~\big(u + s-1 \big)^{2(s-1)}  \label{hoverbetasq}  \,,	
\end{align}
which explicitly vanishes at $u=1$ and diverges at $u=-1$. While the location $u=1-s$ is another root if $s>1$, or another divergent point if $s<1$.
More concretely, we should distinguish the following five situations:
\begin{enumerate}[(I)]
    \item $s<0$: $u=1-s$ is a divergent point. The function $h$ is positive and monotonically increasing if $1<u<1-s$. 
It is natural to expect that the value  $\gamma$ should be in the range $1<\gamma<1-s$, such that $1<u<\gamma$ corresponds to the outside region of the black hole. We plot $h/\beta^2$ for the cases $s=-2$ (blue) and $s=-1$ (yellow) in Fig. \ref{hperbsqximinus};
 \item $0<s<1$: Though $u=1-s$ is still a divergent point, it is located at the left of $u=1$. The function $h$ is positive and monotonically decreasing if $1-s<u<1$.  Then the expected range for $\gamma$ is $1-s<\gamma<1$, such that $\gamma<u<1$ corresponds to the outside region of the black hole. We plot the cases $s=0.1$ (blue) and $s=0.5$ (yellow) in Fig.~\ref{hperbsqxi0to1}; 
 \item $s=1$: In this case, $u=-1$ is the unique divergent point. The expected range for $\gamma$ should be $-1<\gamma<1$, such that $\gamma<u<1$ corresponds to the outside region of the black hole. We plot the corresponding $h(u)/\beta^2$ in Fig.~\ref{hperbsqxi1}; 
 \item $1<s<2$: $u=1-s$ becomes a zero point and $h$ takes local maximum at $u=0$. The expected range for $\gamma$ should be $0<\gamma<1$. This is because the resulting black hole would possess a negative mass when $1-s<\gamma\le 0$. Hence, we require $0<\gamma<1$, such that $\gamma<u<1$  corresponds to the outside region of the black hole. We plot the cases $s=1.5$ (blue) and $s=1.9$ (yellow) in Fig.~\ref{hperbsqxi1to2}; 
\item $s>2$: $u=1-s$ becomes a zero point at the leaf of $u=-1$. Hence we expect $\gamma>1$, such that $h>0$ in $1<u<\gamma$, which corresponds to the outside region of a black hole. We plot the cases  $s=4$ and $s=8$ in Fig.~\ref{hperbsqxi4} and Fig.~\ref{hperbsqxi8}.
\end{enumerate}

The above five situations are summarized in Table~\ref{tab:BumandGaugeViolate} which shows the range of $\gamma$ varies for different coupling $s=\xi/\kappa$ under the attractive gravity condition.

\begin{table}[t]
	\centering  
	\caption{Range from the attractive gravity requirement.
	}  
	\label{tab:BumandGaugeViolate}  
	\begin{tabular}{c|c}  
		\hline \hline
		\hspace{0.3cm} Range of the coupling $\xi$ \hspace{0.3cm} & \hspace{0.3cm} Expected range for $\gamma$ \hspace{0.3cm} \\ [4pt]  
		\hline 
		$\xi< 0 $ & $ 1<\gamma<1-\xi/\kappa  $  \\[7pt]     
		$0<\xi<\kappa$ & $ 1-\xi/\kappa<\gamma <1 $  \\[7pt]
		$\xi=\kappa$ & $-1<\gamma<1 $  \\[7pt]
		$\kappa<\xi<2\kappa$ & $0<\gamma<1 $  \\[7pt] 
		$\xi>2\kappa$ & $\gamma>1 $  \\[7pt]
		\hline
	\end{tabular}
\end{table}

\subsection{Refined range of $\gamma$ from the black hole condition}
\begin{figure}[ht]
	\begin{center}		\includegraphics[width=0.4\textwidth]{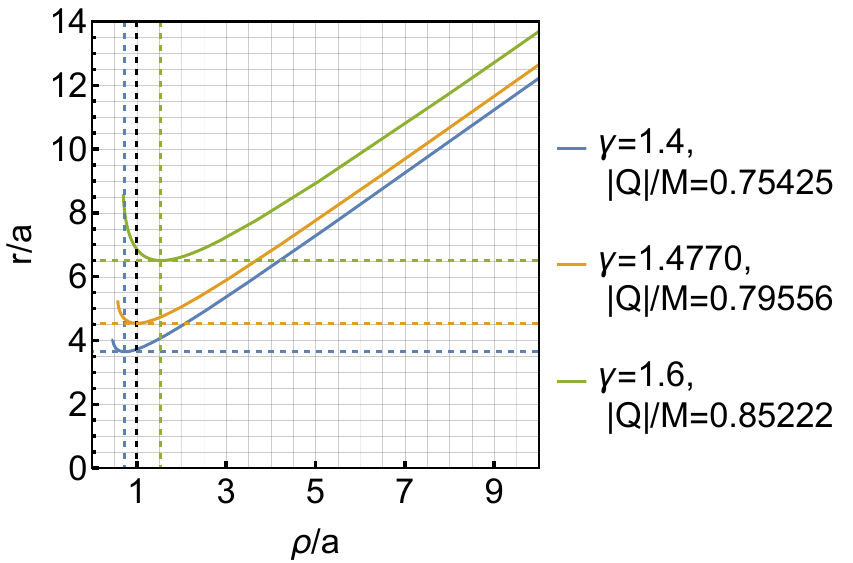}
	\end{center}
	\caption{Functions $r(u)$ for $\xi=-\kappa$.
	 The blue curve corresponds to $\gamma=1.4$ ($|Q|/M=0.75425$) and attains its minimum at $u=0.73321$. This case corresponds to a black bounce solution. The yellow curve depicts the critical case $\gamma=1.4770$ ($|Q|/M=0.79556$), where the minimum of $r$ coincides with the black hole horizon.
	The green curve corresponds to the case $\gamma=1.6$ ($|Q|/M=0.85222$). This case describes a wormhole geometry instead of a black hole, as the minimum of $r$ is located at $u=1.5403$ with a positive $h$. The hypersurface defined by $u=1.5403$ is timelike and thus forms the throat of a traversable wormhole. In all cases, $r$ approaches $\rho$ as $\rho\rightarrow +\infty$.
} 
	\label{throatxim1}  
\end{figure}

We further study the function $r(u)$ to exclude the parameter region that the solution \eqref{BHfam} does not describe a black hole. 
 Since the above section shows that $u=1$ corresponds to the Killing horizon, and $u=\gamma$ corresponds to far regime in the asymptotically flat spacetime, we suspect there should be $1<u<\gamma$ for the case of $\gamma>1$, and $\gamma<u<1$ for the case of $\gamma<1$.
For instance, Fig.~\ref{throatxim1} shows $r(u)$ for the case of $\xi=-\kappa$. Three curves  of $r(u)$ for $\gamma=1.4$, $1.477$, $1.6$ are plotted. They contain the minimum points. Recall that $\rho=a$ (or $u=1$) corresponds to the Killing horizon. 
The minimum point of the curve for $\gamma=1.4$ is behind the horizon. Hence this case corresponds to a black bounce spacetime \cite{Bronnikov:2006fu, Simpson:2018tsi, Huang:2019arj}. It is worth noting that the term ``bounce'' refers to the phenomenon that every infalling geodesic passing through the horizon would experience a subsequent increase in $r$. 
While the critical case is the curve for $\gamma=1.4770$. 
The minimum point for this curve coincides with the horizon $\rho=a$, such that this case also corresponds to a black hole. 
Nevertheless, the minimum point of the curve for $\gamma=1.6$ is outside the horizon. 
Therefore, the minimum point represents a throat for a traversable wormhole connecting an asymptotically flat world and another universe. The Killing horizon $\rho=a$ is more likely a cosmological horizon rather than a black hole horizon.  

Now, we systematically study the extreme points of $r(u)$ to further constrain the range of $\gamma$. 
Since the extreme points for $r$ are also the extreme points for $r^2$,  we focus on $\der (r^2)/\der u$ for convenience. This function is given by 
\begin{widetext}
\begin{align}
 \frac{\der r^2}{\der u}=&~ \frac{2e^{2(u+\gamma)(\gamma-1)^2} u(u+1)}{u+s-1} \bigg|\frac{u+s-1}{\gamma+s-1}\bigg|^{2(1-s)} 
 \frac{ e^{2\gamma} (\gamma-1)(u+1)^2 + e^{2u}(\gamma+1)(u^2+2(s-1)u+1) }{ \big[e^{2\gamma} (\gamma-1)(u+1) - e^{2u}(\gamma+1)(u-1)\big]^3 } \,,
\end{align}
such that we obtain an equation controlling the extreme points of $r^2(u)$, which is
\begin{align}
	0=&~ e^{2\gamma} (\gamma-1)(u_{\text{ex}}+1)^2 
   + e^{2u_{\text{ex}}}(\gamma+1)(u_{\text{ex}}^2+2(s-1)u_{\text{ex}}+1)  \,,\label{eq4extArealrad}
\end{align}
\end{widetext}
where the subscript ``ex'' emphasizes that $u=u_{\text{ex}}$ is the extreme point of $r^2(u)$.
\begin{figure}[h]
	\begin{center}
		\includegraphics[width=0.27\textwidth]{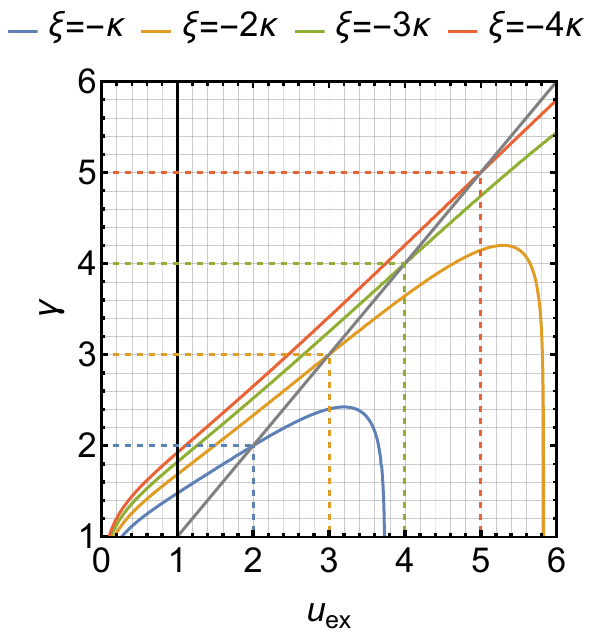}
  \includegraphics[width=0.27\textwidth]{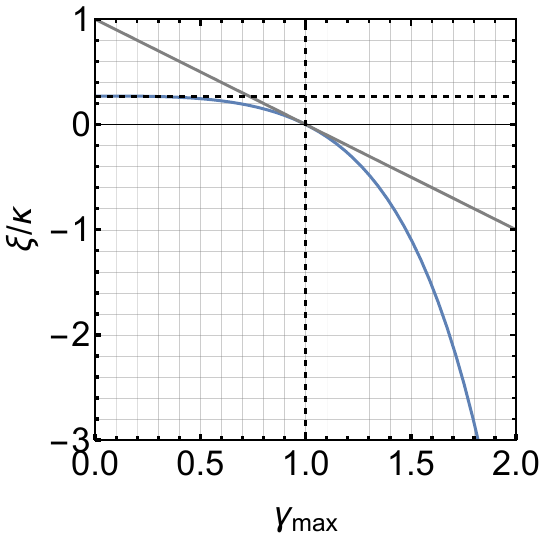}
	\end{center}
	\caption{\small Constraints for black hole solutions when $\xi<0$. Upper panel: extremal points $u_{\text{ex}}$ of the areal radius versus the parameter $\gamma$. Considering the rough range corresponding to the outside of the black hole is $1<u<\gamma$, there should be $u_{\text{ex}}\le 1$ to ensure no traversable wormhole throat appears outside the horizon $u=1$. Hence $\gamma$ should be upper limited. Lower panel: Analytic $\gamma_{\text{max}}$-$\xi/\kappa$ relation, i.e., Eq.~\eqref{gammaxVSs}, which defines the upper bound of $\gamma$ for black hole solutions. Particularly, The $\gamma_{\text{max}}>1$ regime corresponds $\xi<0$. } 
	\label{throatrelmarkxim}  
\end{figure}
The upper plot in Fig.~\ref{throatrelmarkxim} shows that the $\gamma$ range for the case of $s<0$ should be refined as $1<\gamma \le \gamma_{\text{max}}$, where $ \gamma_{\text{max}}$ denotes the maximum value of $\gamma$.
One can verify that its relation with the reduced coupling $s=\xi/\kappa$ is
\begin{align}
	s = 2 e^{2(\gamma_{\text{max}}-1)}\, \frac{1-\gamma_{\text{max}}}{1+\gamma_{\text{max}}} \,, \label{gammaxVSs}
\end{align}
which is derived via substituting $u=1$ into Eq.~\eqref{eq4extArealrad}.
We call such a relation as the $\gamma_{\text{max}}$-$s$ relation, as shown in the lower panel of Fig.~\ref{throatrelmarkxim}.

\begin{figure}[t]
	\begin{center}		
    \includegraphics[width=0.28\textwidth]{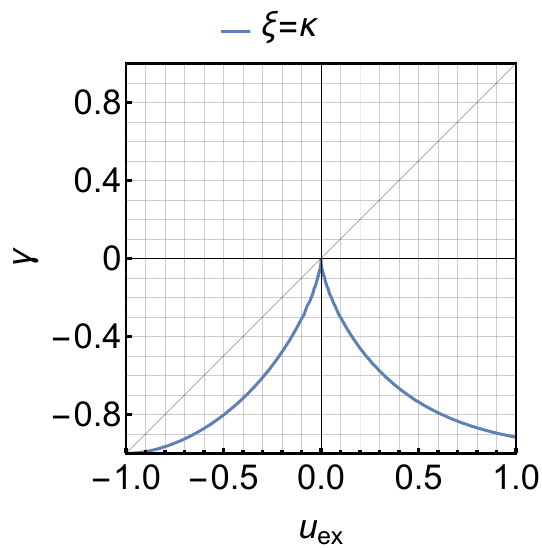}
\includegraphics[width=0.31\textwidth]{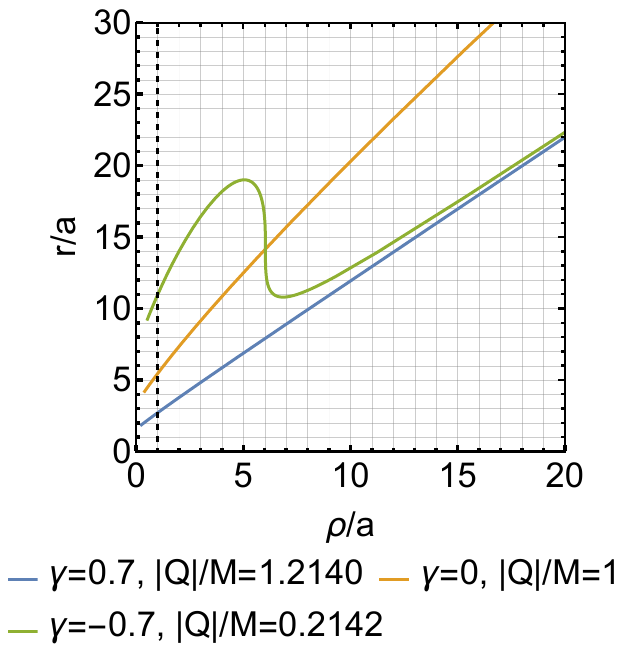}
	\end{center}
	\caption{Case of $\xi=\kappa$. Upper panel: relation between the extremal point $u_{\text{ex}}$ of the areal radius and  $\gamma$, showing no extrema for $\gamma>0$, and two extrema for $\gamma<0$. Lower panel: areal radius $r$ as a function of $\rho$. The case of $\gamma=-0.7$ exhibits a ``Python's lunch'' geometry with a local maximum and minimum of the areal radius. } 
	\label{throatxi1}  
\end{figure}

\begin{figure}[t!]
	\begin{center}
		\includegraphics[width=0.3\textwidth]{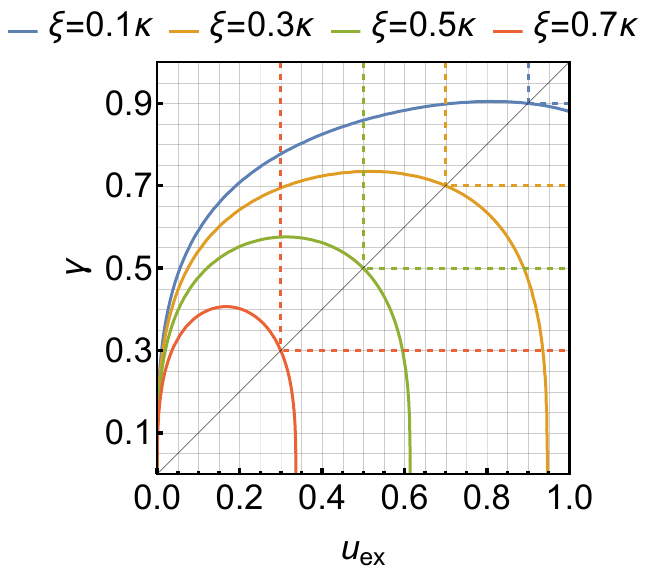}

\includegraphics[width=0.27\textwidth]{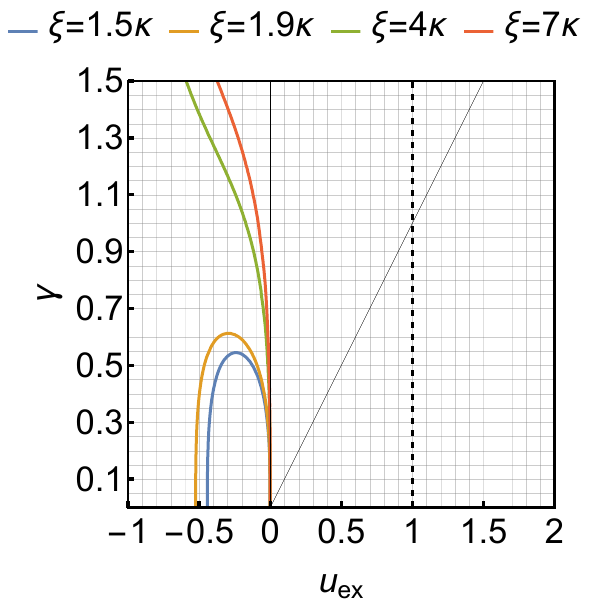}
	\end{center}
	\caption{ 
    Throat location $u_{\text{ex}}$ versus $\gamma$ for $0<\xi<\kappa$ (upper panel) and $\xi>\kappa$ (lower panel). 
    In the case of $0<\xi<\kappa$, the rough range for $u$ is $\gamma<u<1$, so the range of $u_{\text{ex}}$ for the appearance of wormhole throat outside the horizon should be $\gamma<u_{\text{ex}}<1$, the relation curves do not enter the appropriate parameter regime.
    For the the case of $\xi>\kappa$, no $u_{\text{ex}}$ appears in the rough range $u>1$.
    Therefore, there is no additional constraint on $\gamma$ for black hole solutions. } 
	\label{nothroatxi}  
\end{figure}

\begin{figure}[t!]
\begin{center}	\includegraphics[width=0.32\textwidth]{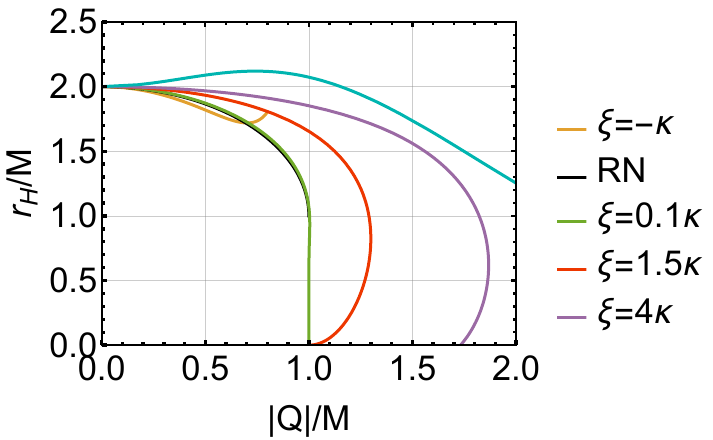}
\newline
\newline
\includegraphics[width=0.26\textwidth]{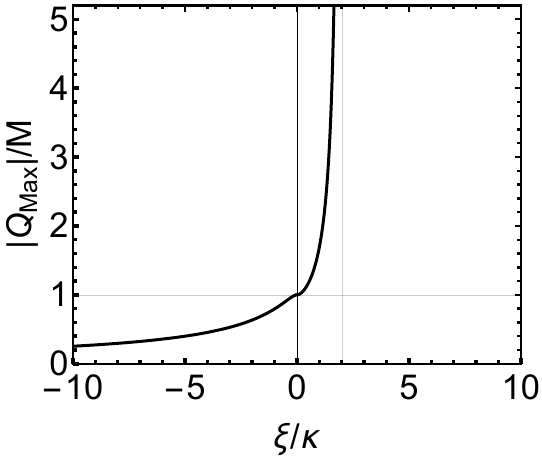}
\end{center}
\caption{\small  Upper panel: horizon radius $r_H$ versus bumblebee charge $|Q|$ in the unit of ADM mass for different coupling values, including the case of Reissner--Nordstr\"om (RN) black hole.
Lower panel: charge-mass ratio $|Q|/M$ as a function of the reduced coupling $s=\xi/\kappa$. We emphasize that there is no upper limit for $\xi>2\kappa$ (i.e., $s\ge 2$). } 
\label{matchrHVSQ}  
\end{figure}

Then, another situation that the solution \eqref{BHfam} does not describe a black hole occurs for the case of $s=1$, as is shown in Fig.~\ref{throatxi1}. 
The upper panel of Fig.~\ref{throatxi1} represents the relation between $u_\text{ex}$ and $\gamma$. Suppose the value of $\gamma$ is taken negative, then a constant line for the constant $\gamma$ would hit the $u_\text{ex}$-$s$ relation curve twice.
For example, the green curve for the case of $\gamma=-0.7$ in the lower plot of Fig.~\ref{throatxi1} indicates that the spatial geometry is not a two-sided traversable wormhole, but a ``Python's lunch'' spatial geometry. Such a configuration contains the local maximum point and the local minimum point, which was proposed in the study of decoding Hawking radiation \cite{Brown:2019rox}.

On the other hand, Fig.~\ref{nothroatxi} shows that there is no additional constraint on $\gamma$ when $0<s<1$ and $s>1$.
The upper plot shows the case of $0<s<1$. Since $u$ includes $u=u_{\text{ex}}$, and the value of $\gamma$ should satisfy $\gamma<u_{\text{ex}}<1$, we conclude that only the region above the diagonal line (gray) corresponds to black holes, while the $u_{\text{ex}}$-$\gamma$ contour is excluded.  
Then, the lower plot of Fig.~\ref{nothroatxi} for the case of $s>1$ shows that they never encounter $u=u_{\text{ex}}$ in the region satisfying the attractive gravity condition (see Table ~\ref{tab:BumandGaugeViolate}). This is because all $u=u_{\text{ex}}$ for these cases locate at the negative region of $u_{\text{ex}}$. 
Finally, we summarize the refined range of $\gamma$ for different $s$ in Table~\ref{tab:BumBHgammaRange}.

\begin{table}[t]
\centering  
\caption{Refined range from no traversable throat condition.}  
\label{tab:BumBHgammaRange}  
\begin{tabular}{c|c}  
	\hline \hline
	\hspace{0.3cm} Range of the coupling $\xi$ \hspace{0.3cm} & \hspace{0.3cm} Refined range of $\gamma$ \hspace{0.3cm}  \\ [4pt]  
	\hline 
	$\xi< 0 $ & $ 1 < \gamma \le \gamma_\text{max}$  \\[7pt]     
	$0<\xi<\kappa$ & $ 1-\xi/\kappa<\gamma <1 $  \\[7pt]	
    $\xi=\kappa$ & $ 0\le\gamma<1 $  \\[7pt]
	$\kappa<\xi < 2\kappa$ & $0<\gamma<1 $  \\[7pt]
	$\xi>2\kappa$ & $\gamma>1 $  \\[7pt]
	\hline
\end{tabular}
\end{table}

\begin{figure}[ht]
\begin{center} 
	\includegraphics[width=0.26\textwidth]{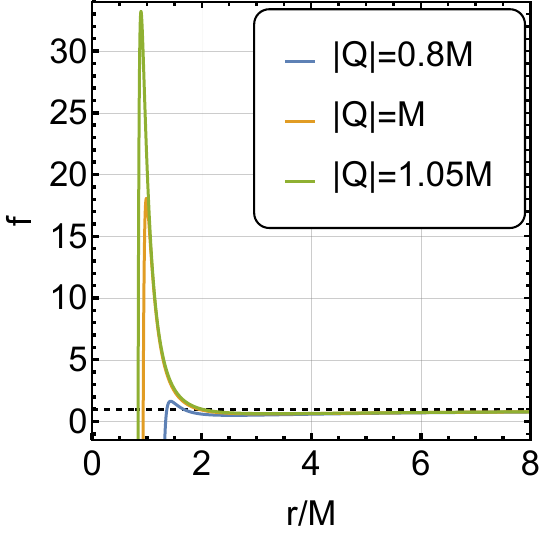}
\end{center}
\caption{\small $f$ versus $r/M$ for $\xi=8\kappa$. When the charge-mass ratio increases, the peak value of the function $f(r)$ before the root also increases. The plot shows the cases of $|Q|=0.8 M$ (blue), $|Q|= M$ (yellow), and $|Q|=1.05 M$ (green). An extremely high peak value is an obstacle to numerical calculation. } 
\label{rVSfxi8}  
\end{figure}

\subsection{Upper bound for $|Q|/M$}
Based on the refined $\gamma$ range, we reinvestigate the $r_H$-$|Q|$ relation, where the unit is set as dimensionless. 
We recover the results in the previous study by \citet{Mai:2023ggs}, which are shown in the upper panel of Fig.~\ref{matchrHVSQ}. 
Meanwhile, the lower panel shows the $s$-$|Q/M|$ relation.
Despite the situation $0<s<2$ where there is indeed a finite maximum $|Q/M|$ from the formula, the maximum charge-mass ratio for the case of $s<0$ arises from the requirement $\gamma\le\gamma_{\text{max}}$, which indicates that a wormhole throat is located outside the Killing horizon.
As for the case of $s>2$, there is no maximum charge-mass ratio since there is no upper bound for $\gamma$. 
This is a feature that was obscured by the uncontrolled numerical errors in early studies \cite{Xu:2022frb, Mai:2023ggs, Mai:2024lgk}.

The peak of $f$ explains why the numerical solution encounters the numerical error.
When $|Q|/M$ becomes large, $f$ develops a peak region located close to, but outside, the horizon. 
For example, Fig.~\ref{rVSfxi8} shows that a small change in $|Q|/M$ leads to a rapid increase in the peak value of $f$. 
Such an extremely high value of $f$ is problematic for controlling the numerical precision. 
This is why earlier studies misidentified the maximum charge-mass ratio for $\xi>2\kappa$.

\section{Thermodynamics} \label{sec:thermo}

In this section, we study black hole thermodynamics.
The analytic expression of the solution \eqref{BHfam} allows us to scan the whole parameter space in a lower cost than the numerical approach~\cite{Mai:2023ggs}. 
As a result, we not only confirm the numerical results of \citet{Mai:2023ggs} with higher precision, but also discover several new cases which were not identified.

\subsection{Entropy and Temperature}

\begin{figure}[ht]
	\begin{center}
	\includegraphics[width=0.34\textwidth]{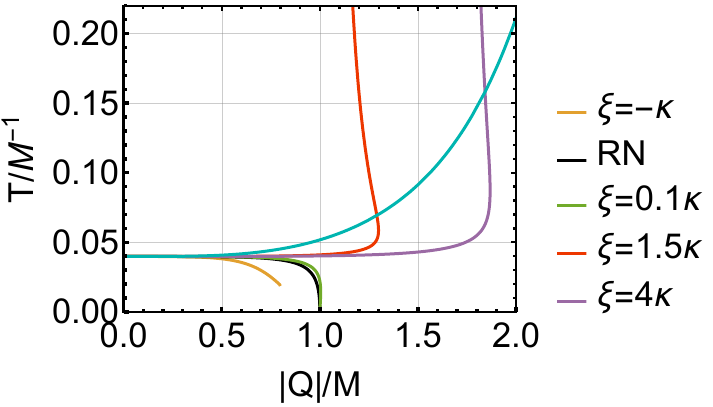} 
	\newline
    \newline
    \newline
\includegraphics[width=0.28\textwidth]{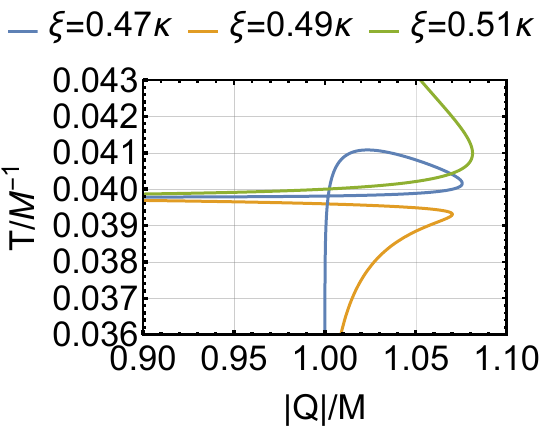} 
	\end{center}
	\caption{Hawking temperature $T$ (in unit of $M^{-1}$) versus the charge-mass ratio $|Q|/M$. Upper panel: results confirming previous numerical investigation~\cite{Mai:2023ggs}. Lower panel: nonmonotonic turning behavior for $\xi=0.49\kappa$, missed in Ref.~\cite{Mai:2023ggs}.  } 
\label{matchTemVSQ}  
\end{figure}

As pointed out by the reference~\cite{An:2024fzf}, the bumblebee vector field $b^{\mu}$ may affect the Wald entropy of the Schwarzschild-like black hole. The relation of entropy and area may be different from the black holes in general relativity. However, this is not the case for the above solutions \eqref{BHfam} in which $b^{\mu}$ only contains the temporal component.
To manifest this property, we express $b^{\mu}$ in terms of the Killing vector $\chi^{\mu}=\partial x^{\mu}/\partial t$ as follows:
\begin{widetext}
\begin{align}
	b^{\mu} =& -\frac{b_t}{h} \frac{\partial x^{\mu}}{\partial t} 
    =-\sqrt{\frac{2}{\kappa}}\, \frac{  u+1 }{  \beta s(s-2)  } \big(u + s-1 \big)^{1-s} \, \chi^{\mu}  \,.
\end{align}
The vector field $b^{\mu}$ vanishes on the bifurcate surface since the Killing vector field $\chi^{\mu}$ vanishes there. Hence the entropy of the bumblebee black hole is still $S= A_H/4$, where $A_H$ denotes the horizon area. Particularly, the entropy $S$ is
\begin{align}
 S  =&~ 4\pi a^2  |s|^{2(1-s)} e^{2(2-\gamma)} \big(\gamma + s-1 \big)^{2(s-1)}
 \, . \label{entropybyagam}
\end{align}

On the other hand, the black hole temperature $T$ is given by $T=\kappa_H/(2\pi)$. Choosing the inverse of the ADM mass $M$ as the unit for temperature, the result of surface gravity $\kappa_H$ given in Section~\ref{sec:theory} yields
\begin{align}
	T M=&~ \frac{M}{4\pi}\frac{\der h}{\der \rho} 
    = \frac{|s|^{2(s-1)} }{8\pi \gamma } e^{2\gamma-2 }  (\gamma-1+s)^{1-2s}[(s-1)\gamma+1] \,.\label{TeminvMasunitbyagam}
\end{align}
\end{widetext}
Then, combining it with the analytic result for the charge-mass ratio in Section~\ref{sec:BHs}, we reproduce the $T$-$|Q|/M$ relations in the upper panel of Fig.~\ref{matchTemVSQ}, which matches the results in Ref.~\cite{Mai:2023ggs}.
Furthermore, we found some missing cases in the previous work. 
As shown in the lower panel of Fig.~\ref{matchTemVSQ}, when $\xi=0.49\kappa$,  the $T$-$|Q|/M$ relation exhibits a turning point, and thus has a degenerate region in which the same $|Q|/M$ corresponds to different temperatures. 
It is distinct from the decreasing case $\xi=0.47\kappa$ and the increasing case $\xi=0.51\kappa$.
Nevertheless, we now observe that the scanning step used in Ref.~\cite{Mai:2023ggs} is too large to cover $\xi=0.49\kappa$.

\subsection{$Y$ charge and $X$ potential}
\begin{figure}[t]
	\begin{center}
		\includegraphics[width=4.1cm]{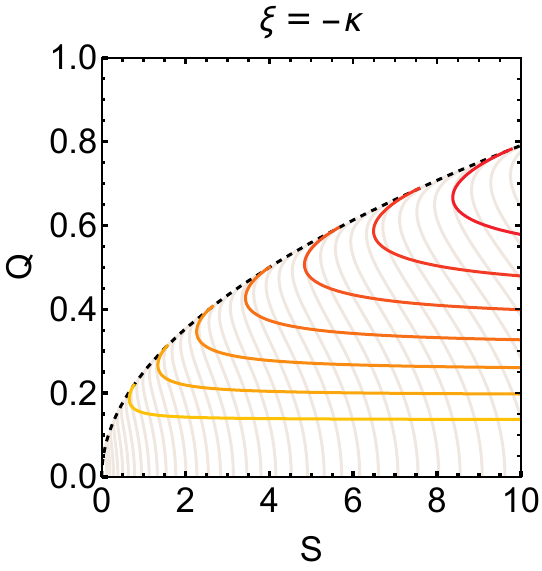}
		\includegraphics[width=4.1cm]{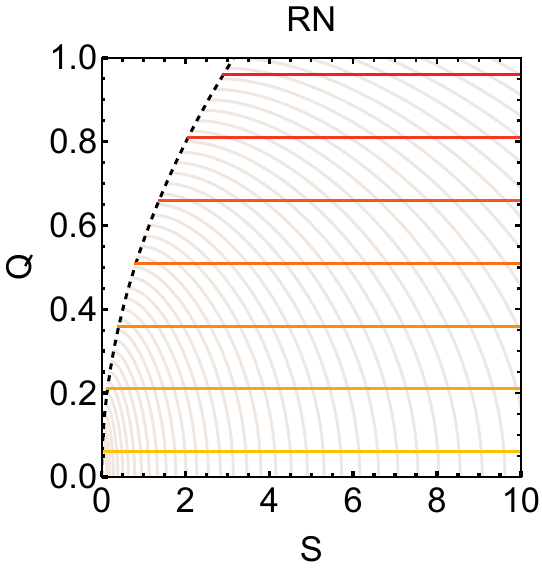} 
		\includegraphics[width=4.1cm]{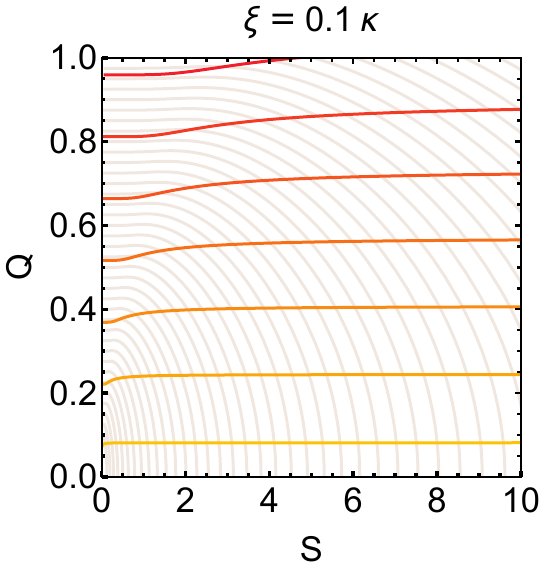}
		\includegraphics[width=4.1cm]{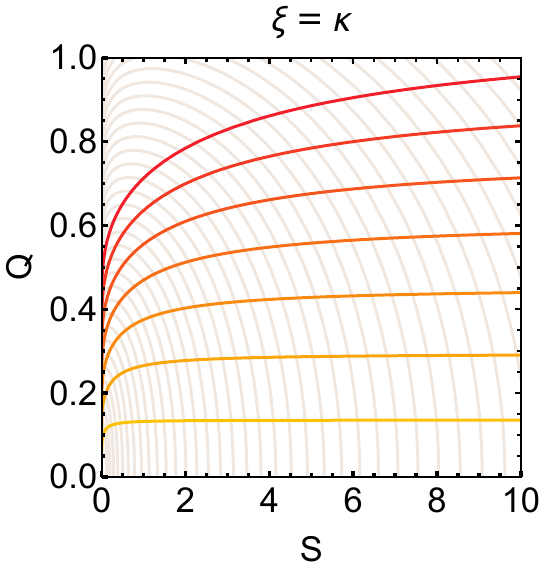}
		\includegraphics[width=4.1cm]{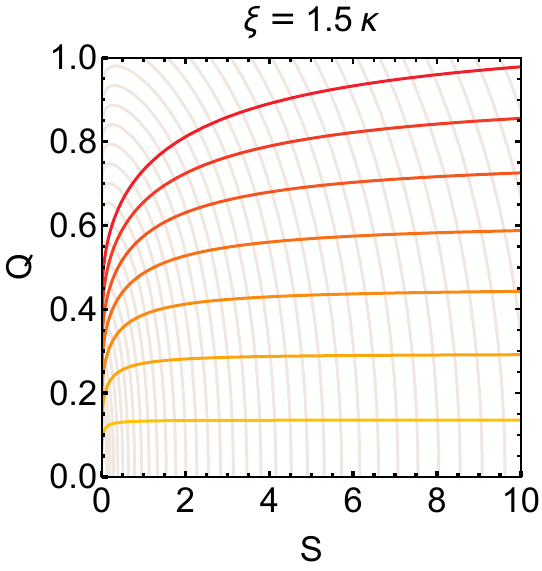}
		\includegraphics[width=4.1cm]{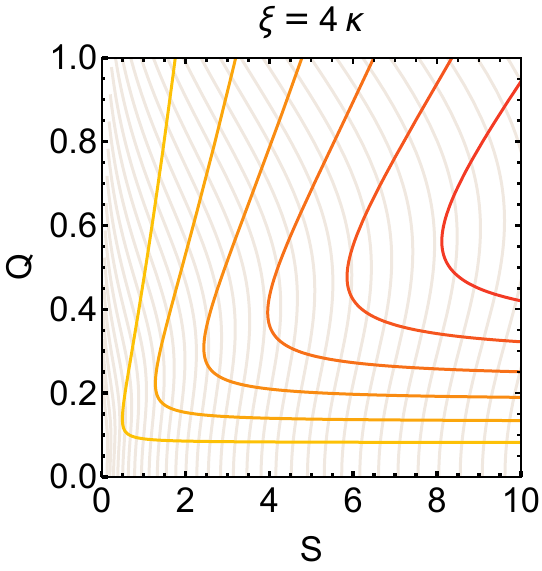}
\end{center}	\caption{\small Contours of constant $Y$ charge in the dimensionless entropy-charge ($S$-$Q$) plane for $\xi=-\kappa$, RN black hole, $\xi=0.1\kappa$, $\xi=\kappa$, $\xi=1.5\kappa$, and $\xi=4\kappa$. Our analytic results reproduce and validate the extended thermodynamic framework established in prior numerical studies~\cite{Mai:2023ggs}. } 
	\label{SvsQ}  
\end{figure}

To avoid issues arising from treating $\{M, S, Q\}$ as the basic thermodynamic variables, we follow the same approach as in Ref.~\cite{Mai:2023ggs} to construct the $Y$ charge and the $X$ potential, which are constructed to restore the first law of thermodynamics and Smarr relation,
\begin{align}
	\delta Y = \frac{ \delta M - T\delta S }{X} \,,\quad 
	M=2TS+XY \,.
\end{align}
More concretely, the $Y$ charge should be a function of solution parameters $\{a,\gamma\}$. 
We require the following two conditions: 
(i) $\delta Y$ is integrable;
(ii) $Y$ has the same dimension as $M$ or $Q$, i.e., $Y$ should be proportional to the dimensional parameter $a$.
These two conditions suggest the following ansatz for constructing $Y$ and $X$:
\begin{align}
	M = a \tilde{M}(\gamma) \,,\quad
	S = a^2 \tilde{S}(\gamma) \,, \quad
	Y = a \tilde{Y}(\gamma) \,,
\end{align}
where the argument $\gamma$ means that functions $\{\tilde{M}(\gamma), \tilde{S}(\gamma), \tilde{Y}(\gamma)\}$ only depend on the variable $\gamma$. 
We emphasize that $\tilde{M}(\gamma)$ and $ \tilde{S}(\gamma)$ can be read from Eq.~\eqref{ADMmassbyagam} and Eq.~\eqref{entropybyagam}.
According to Eq.~\eqref{kappaXrHsq}, there is $T=a/(2S)= (2a\tilde{S})^{-1}$.
Then, we obtain
\begin{align}
	\delta M - T\delta S  =&~ \big(\tilde{M} -1\big) \delta a + a\bigg( \frac{\der\tilde{M}}{\der \gamma} -\frac{1}{2 \tilde{S} } \frac{\der\tilde{S}}{\der \gamma}\bigg) \delta \gamma \, .
\end{align}
If we require the above expression to be equal to $X\delta Y=X\tilde{Y}\delta a + aX \delta\tilde{Y} $, there should be
\begin{align}
	\tilde{M} -1  =&~ X \tilde{Y} \label{preSmarr} \,, \\
	\frac{\der\tilde{M}}{\der \gamma} -\frac{1}{2 \tilde{S} } \frac{\der\tilde{S}}{\der \gamma} =&~ X \frac{\der\tilde{Y}}{\der \gamma} \label{forIntegral}
	\,, 
\end{align}
where the first equation confirms the Smarr relation $M = a +XY =2TS+XY $, and the second one implies that
\begin{align}
	\frac{1}{\tilde{M} -1}\frac{\der(\tilde{M} -1)}{\der \gamma} -\frac{1}{2 \tilde{S} (\tilde{M} -1) } \frac{\der\tilde{S}}{\der \gamma} =&~ \frac{1}{ \tilde{Y} }\frac{\der\tilde{Y}}{\der \gamma} \label{YchargeEq2} \,.
\end{align}
It hints at constructing the analytic $Y$ charge and the $X$ potential up to an overall constant $Y_0$ as follows:
\begin{align}
	Y &= a Y_0~ e^{-\gamma} \frac{\gamma+1}{\gamma+s-1}  \frac{\sqrt{ \big|1-\gamma^2\big| } }{\gamma}
	\,, \nonumber \\
	X &= Y_0^{-1}~ e^{\gamma} \sqrt{ \bigg|\frac{1-\gamma}{1+\gamma}\bigg| }
	\,.\label{anlyXpotentialYcharge}
\end{align}
Substituting Eqs.~\eqref{entropybyagam}, \eqref{TeminvMasunitbyagam}, and \eqref{anlyXpotentialYcharge} into the first law $\delta M=T\delta S+X\delta Y$ confirms its validity for all the parameter regimes.
Based on these results, we redraw those $Y$-contours in the $S$-$Q$ space in Fig.~\ref{SvsQ}, using the same choice of units for $S$, $Q$, and $Y$ as that was used in Ref.~\cite{Mai:2023ggs}. 
Most of the early results are confirmed, and the error that appeared in the case of $\xi=4\kappa$ due to low numerical precision is corrected.

\subsection{Heat capacity}
We then calculate the heat capacity for constant-$Y$ processes, and study its divergent points and vanishing points.
As pointed out in Ref.~\cite{Mai:2023ggs}, the vanishing point of $C_Y$ may imply a smooth phase transition from the locally thermodynamic unstable phase (indicated by $C_Y<0$) to a stable phase (indicated by $C_Y>0$), while the divergent point of $C_Y$ may imply a discontinuous phase transition.

A constant-$Y$ process induces a relation between $a$ and $\gamma$ as follows:
\begin{align}
 a = \frac{Y}{Y_0} \frac{\gamma+s-1}{\gamma+1}   \frac{\gamma e^{\gamma} }{\sqrt{ \big|1-\gamma^2\big| } }
		\,. 
\end{align}
Then we rewrite the ADM mass $M$ and the temperature $T$ as functions of $Y$ and $\gamma$ as:
\begin{align}
	M =&~ \frac{Y}{Y_0} \frac{(s-1)~ \gamma +1}{\gamma+1}   \frac{ e^{\gamma} }{\sqrt{ \big|1-\gamma^2\big| } }   \,, \\
	T
	=&~ \frac{Y_0}{Y} e^{\gamma-2 } \frac{\gamma+1}{8\pi \gamma} \sqrt{ \big|1-\gamma^2\big| } \,|s|^{2(s-1)} (\gamma-1+s)^{1-2s} \,. 	
\end{align} 
The value of the heat capacity for constant-$Y$ processes under the unit $M^2$ is
\begin{widetext}
\begin{align}
  C_Y  M^{-2} =&~ \frac{1}{M^2}\bigg(\frac{\partial M(Y,\gamma)}{\partial \gamma} \bigg)_{Y} ~\bigg/~ \bigg(\frac{\partial T(Y,\gamma)}{\partial \gamma} \bigg)_{Y}  \nonumber\\
  =& \bigg(\frac{\gamma+s-1}{s}\bigg)^{2s} \frac{ 8\pi s^2 \gamma^2 e^{2(1-\gamma)}}{((s-1)\gamma+1)^2 } 
  \frac{(s-1)\gamma^3 - (s-2)\gamma^2 -2\gamma -(s-1)}{\gamma^4 -(s-1)\gamma^3 + (s-3)\gamma^2 +2\gamma +(s-1)}  \,.
\end{align}
\end{widetext}
Moreover, the analytic expression for the heat capacity hints at finding vanishing points and divergent points. 
It seems that $\gamma=1-s$ is the divergent point if $s<0$, or the vanishing point if $s>0$. 
However, as pointed out in Section~\ref{sec:theory}, $\gamma=1-s$ is not in the appropriate range of $\gamma$. 
In addition, though $\gamma=1/(1-s)$ seems to be a divergent point, it also locates outside the appropriate range of $\gamma$.
For the cases of $\xi<0$, $s=\xi/\kappa$ is negative. Hence $1-s>1$, i.e., $1/(1-s)<1$. The apparent divergent point $\gamma=1/(1-s)$ does not lie in the appropriate range $1<\gamma<\gamma_{\text{max}}$; 
For the cases of $0<\xi<\kappa$, $-1<-s<0$ implies $0<1-s<1$, such that $\gamma=1/(1-s)$ is larger than $1$, outside the range $1-\xi/\kappa<\gamma<1$;
For the cases of $\xi>\kappa$ (including the case of $\xi=2\kappa$), $-s<-1$ leads to a negative $1-s$. Thus $\gamma=1/(1-s)$ does not satisfy the positive $\gamma$ requirement. These are summarized in Table~\ref{tab:1over1msOutsideRange}.

\begin{figure}[t]
	\begin{center}
		\includegraphics[width=0.33\textwidth]{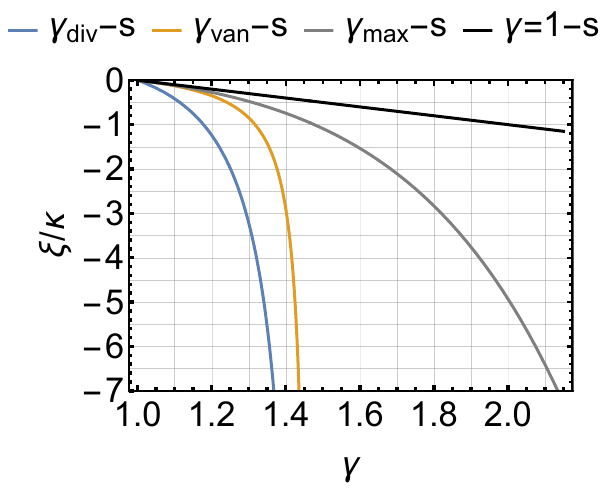}
\newline
\newline
\includegraphics[width=0.35\textwidth]{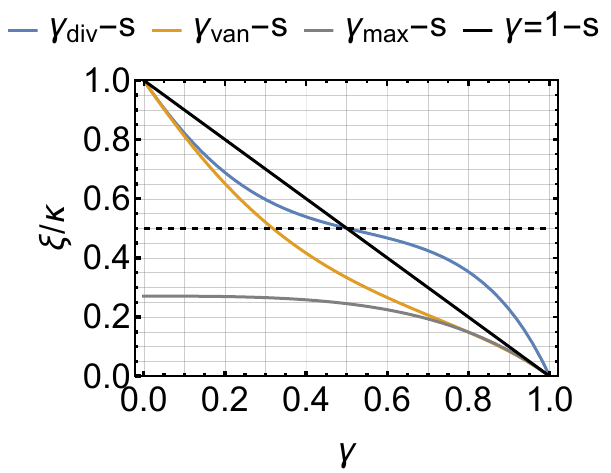}
\newline
\newline
\includegraphics[width=0.29\textwidth]{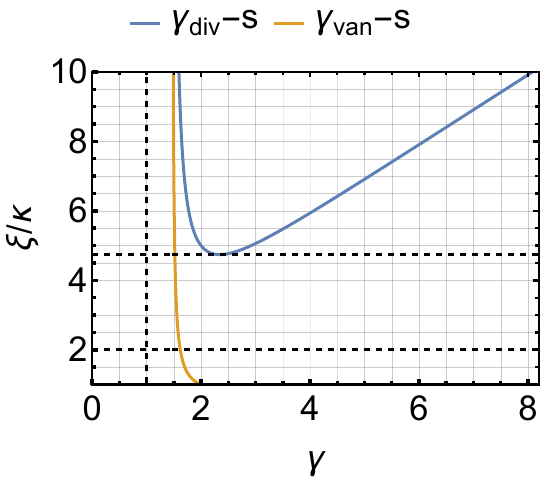}
	\end{center}
\caption{Analytic relations for the vanishing points $\gamma_{\text{van}}$ and divergent points $\gamma_{\text{div}}$ of the constant-$Y$ heat capacity $C_Y$, alongside the black hole boundary $\gamma_{\text{max}}$ . Upper panel: $\xi<0$ regime; Middle panel: $0<\xi<\kappa$ regime; Lower panel: $\xi>\kappa$ regime, showing two divergent points for $\xi>4.74\kappa$. } 
\label{CYcharac}  
\end{figure}

\begin{table}[htbp]  
	\centering  
	  \caption{ The appropriate range of $\gamma$ and the apparent divergent point $\gamma=1/(1-s) \,$.
	}
\label{tab:1over1msOutsideRange}  
	\begin{tabular}{c|c|c}  
		\hline \hline
		&  \\[-8pt]
		\hspace{0.1cm} Range of the coupling $\xi$ \hspace{0.1cm} & Range for $\gamma$  & \hspace{0.1cm} $\gamma=1/(1-s)$ \hspace{0.1cm}  \\ [4pt]  
		\hline 
		&  \\[-9pt]
		$\xi< 0 $ & $ 1 < \gamma < \gamma_\text{max}$  &  $1/(1-s)<1$  \\[7pt]     
		$0<\xi<\kappa$ & $ 1-\xi/\kappa<\gamma <1 $  & $1/(1-s)>1$  \\[7pt]
		$\xi > \kappa$ & $\gamma>0 $ &  $1/(1-s) < 0$  \\[7pt]
		\hline
	\end{tabular}
\end{table}
Therefore, divergent points of $C_Y$ are controlled by 
\begin{align}
	\gamma_{\text{div}}^4 -(s-1)\gamma_{\text{div}}^3 + (s-3)\gamma_{\text{div}}^2 +2\gamma_{\text{div}} +(s-1)=0 \,,
\end{align}
where the subscript ``$\text{div}$'' denotes the divergent point, while vanishing points of $C_Y$ is controlled by 
\begin{align}
	(s-1)\gamma_{\text{van}}^3 - (s-2)\gamma_{\text{van}}^2 -2\gamma_{\text{van}} -(s-1)=0 \,,
\end{align}
where the subscript ``$\text{van}$'' denotes the vanishing point. The solutions of the above equations yield two relations between $s$ and $\gamma$.  
Then we obtain these two relation curves in the $\gamma$-$s$ space:
\begin{align}
	&\gamma_{\text{div}}\mbox{-}s~\text{ relation}:~ s= \frac{\gamma_{\text{div}}^4+\gamma_{\text{div}}^3-3\gamma_{\text{div}}^2+2\gamma_{\text{div}}-1}{ \gamma_{\text{div}}^3 - \gamma_{\text{div}}^2 -1 } \,,\nonumber\\
	&\gamma_{\text{van}}\mbox{-}s~\text{ relation}:~ s= \frac{\gamma_{\text{van}}^3-2\gamma_{\text{van}}^2 +2\gamma_{\text{van}} -1 }{ \gamma_{\text{van}}^3 - \gamma_{\text{van}}^2 -1 } 	\,.
\end{align}
We plot these curves and the relation \eqref{gammaxVSs} in Fig.~\ref{CYcharac}. 
The upper panel of Fig.~\ref{CYcharac} shows the case of $s<0$. If we fix the value of $s$, represented by a horizontal line, this line intersects $\gamma_{\text{div}}-s$,  $\gamma_{\text{van}} - s$, and $\gamma_{\text{max}} - s$. 
The upper panel shows that $1<\gamma_{\text{div}}<\gamma_{\text{van}}<\gamma=\gamma_{\text{van}}$. Recall that the range $1<\gamma<\gamma_{\text{max}}$ corresponds to a black hole. 
As $\gamma$ increases, one would firstly encounter the divergent $C_Y$ at $\gamma=\gamma_{\text{div}}$, then meet $C_Y=0$ at $\gamma=\gamma_{\text{van}}$. The black line $\gamma=1-s$ is irrelevant for cases of $s<0$, since it lies far to the right of $\gamma_{\text{max}}$.
However, $\gamma=1-s$ is important for cases of $0<s<1$, in which the range of $\gamma$ should be $1-s<\gamma<1$. Hence, the middle panel of Fig.~\ref{CYcharac} shows that $C_Y$ exhibits one divergence for $0<s<0.5$, while it does not cross zero for $0.5<s<1$. 
While the lower panel of Fig.~\ref{CYcharac} shows that $C_Y$ consists of two branches with a different range of $\gamma$. For $1<s<2$, there is no divergent point or vanishing point of $C_Y$ in $0<\gamma<1$, and thus, $C_Y$ does not change its sign in these cases.
For $s>2$, $C_Y$ generally vanishes once. Nevertheless, $C_Y$ remains finite up to approximately $s\approx 4.74$. Then any $s>4.74$ leads to the situation where $C_Y$ diverges twice, which is not covered in Ref.~\cite{Mai:2023ggs}. 

\begin{figure}[t]
	\begin{center}
		\includegraphics[width=0.41\textwidth]{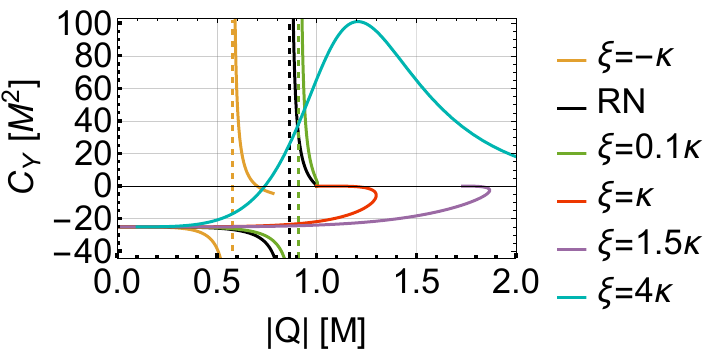}
\newline
\newline
\newline
\includegraphics[width=0.34\textwidth]{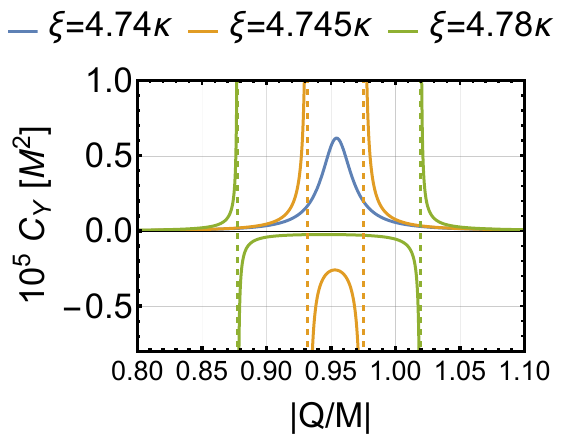}
\newline
\newline
\newline
\includegraphics[width=0.34\textwidth]{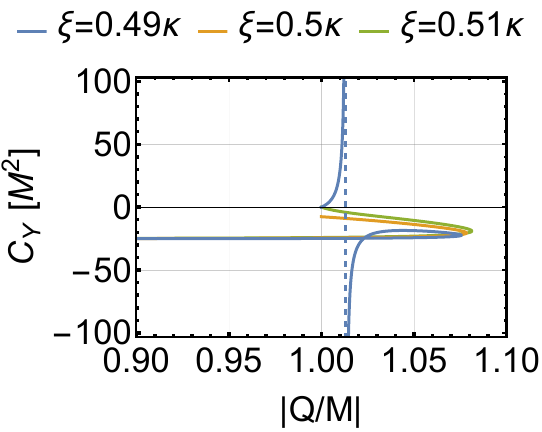}
	\end{center}
	\caption{\small Dimensionless constant-$Y$ heat capacity $C_Y M^{-2}$ versus the charge-mass ratio $|Q|/M$. Upper panel: heat capacity profiles for reproducing prior numerical results;
    Middle panel: double divergence of $C_Y$ for $\xi>4.74\kappa$, a feature missed in Ref.~\cite{Mai:2023ggs}; Lower panel: turning behavior of $C_Y$ for $\xi=0.49\kappa$, matching the nonmonotonic temperature profile in Fig.~\ref{matchTemVSQ} } 
	\label{matchrHCYVSQ}  
\end{figure}

The above discussion classifies all situations where $C_Y$ diverges or vanishes and allows us to study the $C_Y$-$|Q|$ relation in higher precision.  
We reproduce the same results in Ref.~\cite{Mai:2023ggs} in the upper panel of Fig.~\ref{matchrHCYVSQ} and extend these $C_Y$-$|Q/M|$ curves into a larger parameter regime. 
Meanwhile, the middle panel of Fig.~\ref{matchrHCYVSQ} includes new situations with $\xi=4.745\kappa$ and $\xi=4.78\kappa$ where $C_Y$ contains two divergent points, but $C_Y$ remains finite for the case of $\xi=4.74\kappa$. 
Finally, the lower panel shows the situation for $\xi$ near $0.5\kappa$. An interesting feature is that $C_Y$ also exhibits a turning point as $|Q/M|$ increases for the case of $\xi=0.49\kappa$.

\section{Conclusion} \label{sec:con}

We construct the first exact, analytic, asymptotically flat black hole solutions \eqref{BHfam} in static spherically symmetric spacetime for the bumblebee gravity with a purely temporal bumblebee field. 
We focus on systematic parameter space classification and thermodynamic analysis in this study, but leave studies of a full uniqueness theorem, global causal structure analysis, and critical phenomena to future work.

To accommodate the non-trivial analytic structures of these solutions, a systematic approach has been further developed to investigate their spacetime geometry, with which the asymptotic flatness is confirmed and the parameter space regime corresponding to black hole solutions \eqref{BHfam} is constrained.
The parameter regime yielding black holes can be categorized into five distinct cases, depending on the value of the non-minimal coupling $\xi$. This refined classification is summarized in Table \ref{tab:BumBHgammaRange}.
Utilizing these new analytic results, explicit expressions for the charge-mass ratio, black hole temperature, and $Y$ charge have been derived. 
We rigorously show that the charge-mass ratio is unbounded for $\xi>2\kappa$. The reason why numerical studies failed in this regime is that accumulated numerical errors are incurred when evaluating the function $f$, as illustrated in Fig.~\ref{rVSfxi8}. 

The thermodynamics of bumblebee black holes have also been re-investigated. Most results from the prior work \cite{Mai:2023ggs} are confirmed, including the successful introduction of the $Y$ charge and $X$ potential via the requirements of integrability and the Smarr relation, as well as the analysis of the heat capacity $C_Y$ for constant-$Y$ processes.
Moreover, previously unreported cases have been identified. For $\xi=0.49\kappa$, the temperature $T_H$ and the capacity $C_Y$ profiles exhibit a turning behavior as $|Q/M|$ approaches unity. For $\xi$ greater than approximately $4.74\kappa$, $C_Y$ diverges twice.

Furthermore, these analytic solutions, including the additional exact static bumblebee solutions where the VEV $b_{\mu}$ possesses only a temporal component (see Appendix~\ref{appendix}), provide the rigorous analytic foundation for ongoing work, including detailed investigations of the quasinormal modes of asymptotically flat bumblebee black holes and wormholes, a key topic for connecting theoretical predictions with observational data.
We emphasize that the coexistence of black holes and wormholes within the same static solution \eqref{BHfam} suggests that there may exist dynamical solutions in bumblebee theory describing black hole--wormhole conversion \cite{Hayward:1998pp,Shinkai:2002gv,Simpson:2019cer, Yang:2021diz}, a promising direction for future work.
For instance, when $\xi<0$, an excessively large value of $\gamma$ leads to a spacetime geometry containing a traversable wormhole. In addition, the spatial geometry which resembles the ``Python's lunch'' type (see Ref.~\cite{Brown:2019rox}) appears when $\xi=\kappa$ and $-1<\gamma<0$.
Moreover, other exact solutions presented in Appendix~\ref{appendix} also seem to contain situations of wormhole geometry.
Such wormhole configurations that appeared in the vacuum solutions were unrecognized in previous studies.
Further exploration of their spacetime geometry and the derivation of observational signatures will be a valuable direction.

\section*{Acknowledgments} 
We thank Rui Xu and Lirui Yang for useful discussions. 
J. Yang is supported by the Guangzhou Municipal Postdoctoral Research Project Funding, the National Natural Science Foundation of China (Grant No. 11873025, 12133004).
Z.-F. Mai is supported by the ``Hangji'' Action Plan (Guangxi Basic Research Program, Grand No.2026GXNSFBA00640240), the Guangxi Science and Technology Innovation Platform Program (Leitai Action Plan, Grant No.Guike LT2600640026) and the "Guangxi Highland of Innovation Talents'' Program.
D. Liang was supported by the National Natural Science Foundation of China (12405065, 12465013).
L.\ Shao was supported by  the Beijing Natural
Science Foundation (1242018),  the National Natural Science Foundation
of China (12573042),  the Max Planck Partner Group
Program funded by the Max Planck Society and the High-performance
Computing Platform of Peking University.

\begin{widetext}
\appendix
\section{ Construction of Solutions}
\label{appendix}

This appendix provides a brief introduction for constructing the solution \eqref{BHfam} and other exact solutions for Eq. \eqref{odegratt}, Eq. \eqref{odegraxx}, and Eq. \eqref{odegraang}.
A crucial step involves summing Eqs.~\eqref{odegraxx} and \eqref{odegraang} to obtain:
\be
2 = r^2 h'' + 4r r' h'  + 2h\big({r'}^2 + r r''\big) = \fft{\der^2}{\der \rho^2} \big( r^2 h \big) \,,
\ee
which can be directly solved by
\be
r^2h= (\rho-\rho_0)^2 + \alpha   \,. \label{keyconstr}
\ee 
Furthermore, the divergence-free condition $\nabla^{\mu}T_{\mu\nu}=0$ implies that Eq.~\eqref{odegratt} can be derived from an algebraic combination of Eqs. \eqref{odegraang} and \eqref{fromeomvec}, and the derivative of Eq.~\eqref{odegraxx}.
Consequently, we only need to solve the following two equations: 
\be
2\kappa  {b_t'}^2 + \xi \fft{{b_t}^2}{h^2}  {h'}^2 -\fft{h'}{h}\big(2\xi b_t b_t' + h'\big)=\fft{4\alpha h}{\big((\rho-\rho_0)^2 + \alpha\big)^2} 
\label{GravityShsolved}  \,,
\ee
\be
h \fft{\der}{\der \rho} \bigg(\fft{(\rho-\rho_0)^2 + \alpha }{h}\,b_t'\bigg)= \fft{\xi}{2\kappa}\, b_t \fft{\der}{\der \rho} \bigg(\fft{(\rho-\rho_0)^2 + \alpha }{h}\, h'\bigg) \label{VecShsolved} \,.
\ee
The only parameter that matters is $\alpha$ since $\rho_0$ can be eliminated by shifting the coordinate $\rho$, i.e., $\rho \to \rho+\rho_0$. We then construct solutions for Eq. \eqref{GravityShsolved} and Eq. \eqref{VecShsolved} by simply setting $\rho_0=0$. 
\subsection{Extreme family}
We first consider the case of a constant $g^{\mu\nu}b_{\mu}b_{\nu}$, namely,
\be
g^{\mu\nu}b_{\mu}b_{\nu} = -\frac{b_t^2}{h} = -b^2 \,,
\ee 
where $b$ is a constant.
Then Eq. \eqref{GravityShsolved} gives
\be
(b^2 \kappa -2) \bigg( \frac{\der b_t}{\der \rho} \bigg)^2 = \frac{2\alpha b_t^2 }{(\rho^2+\alpha)^2}   \,, \label{extConstr}
\ee
and Eq. \eqref{VecShsolved} becomes
\be
(\xi-2\kappa)\frac{\rho^2+\alpha}{b_t} \bigg( \frac{\der b_t}{\der \rho} \bigg)^2 
= (\xi-\kappa)\frac{\der}{\der \rho}\bigg( (\rho^2+\alpha) \frac{\der b_t}{\der \rho} \bigg) \,. \label{VecEqbconst}
\ee
The unique set of consistent parameter values is
\be
b^2 = \fft{2}{\kappa} \,,\quad \alpha=0 \,.\label{extOnly}
\ee
The reason is as follows. If $b^2 \kappa \neq 2$ and $\alpha\neq 0$, Eq. \eqref{extConstr} implies
\be
\frac{\der b_t}{\der \rho} = \frac{\lambda \, b_t }{\rho^2+\alpha} \,,\label{bconstRes}
\ee
where $\lambda$ should be a non-zero constant satisfying
\be
(b^2 \kappa -2)\lambda^2 = 2\alpha \,.
\ee
However, substituting Eq. \eqref{bconstRes} into Eq. \eqref{VecEqbconst}  yields
\be
(\xi-2\kappa) \frac{\lambda^2 b_t}{\rho^2+\alpha} 
= (\xi-\kappa)\lambda \frac{db_t}{d\rho}
= (\xi-\kappa) \frac{\lambda^2 \, b_t }{\rho^2+\alpha}\,, \label{VecEqbconstRes}
\ee
which leads to a contradiction, as $\kappa b_t/(\rho^2+\alpha)\neq 0$ and $\lambda\neq 0$. 
Conversely, if $b^2\kappa=2$, then $\alpha=0$, and vice versa.
Therefore, the only consistent parameter values are those given in Eqs.~\eqref{extOnly}, which automatically satisfy Eq.~\eqref{extConstr}.
Moreover, when $\alpha=0$, Eq.~\eqref{VecEqbconst} becomes
\be
(s-2)\frac{\rho^2}{b_t} \bigg( \frac{\der b_t}{\der \rho} \bigg)^2 
= (s-1)\frac{\der}{\der\rho}\bigg( \rho^2\frac{\der b_t}{\der\rho} \bigg) \,, \label{extVecEq}
\ee
where $s=\xi/\kappa$ for short.
When $s=1$, Eq.~\eqref{extVecEq} gives a trivial solution: a constant $b_t$ and a Minkowski spacetime up to some coordinate transformation. For $s\neq 1$, there is
\be
\frac{s-2}{s-1}\bigg( \rho^2\frac{\der b_t}{\der\rho} \bigg)  
= b_t\frac{\der}{\der b_t}\bigg( \rho^2\frac{\der b_t}{\der\rho} \bigg) \,.
\ee
Integrating the equation, we have
\be
\frac{s-2}{s-1}\log\bigg( \frac{b_t}{\beta} \bigg) 
= \log\bigg( \frac{\rho^2}{\beta l}\frac{\der b_t}{\der\rho} \bigg) \,,
\ee
where we introduced $\beta$ and $l$ to ensure the argument of
the logarithm is dimensionless.
The parameter $\beta$ also acts as an integration constant.
Treating $1/\rho$ as a function of $b_t$, this equation becomes integrable, such that
\begin{align}
	l\frac{\der \rho}{\rho^2}= \bigg( \frac{b_t}{\beta} \bigg) ^{-\frac{s-2}{s-1} } \frac{\der b_t}{\beta} \;
	\rightarrow \; C - \frac{l}{\rho} &=(s-1) \, \bigg( \frac{b_t}{\beta} \bigg)^{\frac{1}{s-1} } \,.
\end{align}
Therefore, we obtain the following solutions for $b_t(\rho)$ and $h(\rho)$:
\begin{align}
& b_t(\rho)= \beta \bigg(\fft{C}{s-1} + \fft{l}{1-s}\fft{1}{\rho}\bigg)^{s-1} \, ,  \nonumber \\
& h(\rho)= \fft{\kappa}{2}\beta^2 \bigg(\fft{C}{s-1} + \fft{l}{1-s}\fft{1}{\rho}\bigg)^{2(s-1)}
\,.
\end{align}
If $C=0$, the function $h$ does not tend to a constant as $\rho\rightarrow +\infty$, resulting in a metric that is not asymptotically flat. To ensure asymptotic flatness, we must impose the additional condition $C\neq 0$. After rescaling $t$ and $\rho$, we obtain the following expressions
\be
b_t(\rho)=\pm \sqrt{\fft{2}{\kappa}} \bigg(1 + \fft{\rho_c}{\rho}\bigg)^{s-1}
\,,\quad
h(\rho)= \bigg(1 + \fft{\rho_c}{\rho}\bigg)^{2(s-1)}  \,. \label{extsol}
\ee
We name this solution as the extreme family since it reduces to the extreme Reissner--Nordstr\"om (RN) solution when $\xi=0$. The same solution was also constructed in Ref.~\cite{Li:2025rjv, Zhu:2025fiy}.
\subsection{Integrable family}
The above result indicates that the case of a constant $b^2$ can only be achieved when $\alpha=0$ and $b^2=2/\kappa$. Thus, we expect that the cases of $\alpha\neq 0$ correspond to a misalignment between $b_{\mu}b^{\mu}$ and $2/\kappa$. It is natural to introduce a function $w(\rho)$ to evaluate this misalignment:
\begin{equation}
	w(\rho)= \frac{\kappa}{2}\frac{b_t^2(\rho)}{h(\rho)} -1\,.
\end{equation}
Then we will use $\{w(\rho),b_t(\rho)\}$ rather than $\{h(\rho),b_t(\rho)\}$ to construct analytic solutions.
Such a change makes Eq.~\eqref{GravityShsolved} become
\begin{align}
	&~ (1+w)\bigg( \frac{1}{b_t}\frac{\der b_t}{\der\rho} \bigg)^2
	+2s(1+w)\bigg( \frac{1}{2(1+w)} \frac{\der w}{\der\rho} - \frac{1}{b_t}\frac{\der b_t}{\der\rho} \bigg)^2 -\frac{\alpha}{(\rho^2+\alpha)^2} \nonumber
	\\=&~\bigg( \frac{1}{b_t} \frac{\der b_t}{\der\rho} -\frac{1}{2(1+w)}\frac{\der w}{\der\rho} \bigg) 
	\bigg((1+2s+2sw)  \frac{1}{b_t}\frac{\der b_t}{\der\rho} - \frac{1}{ 2(1+w)} \frac{\der w}{\der\rho} \bigg)
	\,, 
\end{align}
and
\be
w\bigg( \frac{1}{b_t}\frac{\der b_t}{\der\rho} \bigg)^2 + \frac{s-1+sw}{1+w} \frac{\der w}{\der\rho}\bigg( \frac{1}{b_t}\frac{\der b_t}{\der\rho} \bigg)
= \frac{\alpha}{(\rho^2+\alpha)^2} + \frac{2s-1+2sw}{4(1+w)^2} \bigg(\frac{\der w}{\der\rho} \bigg)^2
\,, \label{alphaGraConstr}
\ee
which can be treated as a ``separable'' equation for $b_t^{-1}\der b_t/\der \rho$. It contains the following solutions
\begin{align}
	\frac{1}{b_t} \frac{\der b_t}{\der \rho} 	=&~ \frac{s-1+sw}{2w(1+w)}\frac{\der w}{\der \rho}  + \frac{\der\psi}{\der\rho}  \,,
\end{align}
where
\begin{align}
	\frac{\der\psi}{\der\rho} 
	=\pm \frac{1}{\rho^2+\alpha}\sqrt{ \frac{\alpha}{w} +\frac{(s-1)^2 + s(s-2)w }{4w^2(1+w)} \bigg((\rho^2+\alpha)\frac{\der w}{\der\rho}\bigg)^2 } \,. \label{dpsidwOrg} 
\end{align}
Comparing to the discussion in the previous section, $w(\rho)$ is not a constant. Hence one can treat $w$ as an alternative coordinate to hide the factor $(\rho^2+\alpha)$ as follows
\be
\frac{\der z}{\der w}=  \frac{1}{\rho^2+\alpha}\frac{\der\rho}{\der w} \,,\label{dzanddx}
\ee
such that Eq.~\eqref{dpsidwOrg} can be reformulated as
\begin{align}
	\bigg(\frac{d\psi}{dw}\bigg)^2 
	=\frac{\alpha}{w} \bigg( \frac{dz}{dw} \bigg)^2 +\frac{(s-1)^2 + s(s-2)w }{4w^2(1+w)}  \,. \label{solveConstr}
\end{align}
We adjust the ansatz and coordinate
\begin{align}
	&h =  w^{s-1}  \exp\big({2\psi} \big) \,,\nonumber \\
	&b_t = \sqrt{ \frac{2} {\kappa}  } w^{\frac{s-1}{2} } \big(1+w\big)^{\frac{1}{2}} \exp\big({\psi} \big)  \,. \label{altansatzfunctions}
\end{align}
The line-element becomes 
\begin{equation}
	\der s^2=-h(w)dt^2+\fft{1}{h} \bigg(\fft{\der\rho(w)}{\der w}\bigg)^2 \der w^2 +\frac{\rho^2(w)+\alpha}{h(w)}\der\Omega^2  \,.
\end{equation}
Notice that Eq. \eqref{GravityShsolved} is solved by Eq. \eqref{solveConstr}.
While Eq. \eqref{VecShsolved} gives
\begin{align}
	&~0=2w(1+w)\frac{\der^2z}{\der w^2} \bigg((s-1)^2 + s(s-2)w + 2(s-1)w(1+w)\frac{\der\psi}{\der w} \bigg) \nonumber
	\\& + \frac{\der z}{\der w}\bigg( (s-1)^2(1+2w) +s(s-2)w^2-4(s-1)w(1+w)^2\frac{\der\psi}{\der w}  -4w^2(1+w)^2\bigg(\frac{\der\psi}{\der w}\bigg)^2 -4(s-1) w^2(1+w)^2\frac{\der^2\psi}{\der w^2} \bigg) \,. \label{Intfam_Eqs_Sec}
\end{align}
Applying Eq. \eqref{solveConstr} again, i.e., substituting $\der\psi/\der w$ into Eq.~\eqref{Intfam_Eqs_Sec}, a tedious calculation yields the following equation for $\der z/\der w$:
\be
\frac{\der z}{\der w}-4\alpha (1+w)^2 \bigg(\frac{\der z}{\der w}\bigg)^3
+2(1+w)[1+s(s-2)(1+w)]\frac{\der^2 z}{\der w^2}=0 \,.
\ee
Its solution is
\be
\bigg( \frac{\der z}{\der w}\bigg)^2 =\frac{1}{4\alpha} \frac{ s(s-2)w +(s-1)^2}{(1+w)(w_0- w)}  \,, \label{solforzofw}
\ee
where $w_0$ is the integral constant. 
The solution \eqref{solforzofw} indicates that $\rho$ as a function of $w$ should be determined by
\begin{equation}
	\int\frac{\der\rho}{\rho^2+\alpha} = \pm \int \sqrt{\frac{1}{4\alpha} \frac{ s(s-2)w +(s-1)^2}{(1+w)(w_0- w)} } ~ \der w \,.
\end{equation}
The result \eqref{solforzofw} also implies 
\be
\bigg(\frac{\der\psi}{\der w}\bigg)^2 
= \frac{ s(s-2)w +(s-1)^2}{4w(1+w)} \bigg(\frac{1}{w_0- w} + \frac{1}{w}\bigg)
=\frac{ \alpha w_0 }{w^2}  \bigg(\frac{\der z}{\der w}\bigg)^2
\,, \label{solforpsiofw}
\ee
which yields the following integral
\begin{equation}
	\psi = \psi_0\pm \int \frac{1}{2w}  \,\sqrt{\frac{ w_0 }{w_0-w} \frac{s(s-2)w +(s-1)^2 }{1+w} } ~\der w \,. \label{integrate4psi}
\end{equation}
In the following three situations, the above integrals lead to exact solutions which can be written in terms of elementary functions: 
\begin{enumerate}[(I)]
    \item $w_0=-1$;
    \item $w_0=-(s-1)^2/\big(s(s-2)\big) $;
    \item $s=2$ which gives the same $z(w)$ and $\psi(w)$ with $s=0$.
\end{enumerate}
\subsection*{(I) The case of $w_0=-1$ }
When $w_0=-1$, Eq.~\eqref{solforzofw} and Eq.~\eqref{solforpsiofw} become
\begin{align}
	\frac{\der z}{\der w}&=\pm\frac{1}{2(1+w)} \sqrt{-\frac{(s-1)^2 + s(s-2)w }{\alpha} } \,,\\ 
	\frac{\der\psi}{\der w}&=\pm\frac{1}{2w(1+w)} \sqrt{(s-1)^2+s(s-2)w}  \,.
\end{align}
To be consistent, $\alpha$ must be negative. Hence, we introduce $\alpha=-a^2$. Integrating the above formulas directly, we obtain
\begin{align}
	z=&~ z_0 \pm \frac{1}{a} \bigg(u +\frac{1}{2}\log\bigg| \frac{u-1}{u+1}\bigg| \bigg) \,, \label{w0m1z} \\ 
	\psi=&~\psi_0 \pm \frac{1}{2}\bigg( (s-1)\log\bigg|\frac{u-(s-1)}{u+(s-1)} \bigg| - \log\bigg|\frac{u-1}{u+1} \bigg|  \bigg)\,,  \label{w0m1psi}
\end{align}
where
\be
u= \sqrt{(s-1)^2 + s(s-2)w} \,.  \nonumber
\ee

Then we should construct functions $\{h(u), b_t(u), r^2(u)\}$ according to the above results. 
Firstly, $\psi(u)$ and $u(w)$ indicate 
\begin{align}
	h=&~ s(s-2)\beta^2 \bigg( u \mp (s-1) \bigg)^{2(s-1)} \frac{u\mp 1}{u\pm 1}
	\,, \nonumber\\
	b_t=&~ \beta \sqrt{ \frac{2}{\kappa} } (u\mp 1) \bigg( u \mp (s-1) \bigg)^{s-1}\,, \label{hbtbyu}
\end{align}
in which the ``$-$'' sign corresponds to the ``$+$'' branch of Eq. \eqref{w0m1psi}, while the ``$+$'' case of Eq. \eqref{hbtbyu} corresponds to the ``$-$'' branch of Eq.~\eqref{w0m1psi}. 
Next, we will focus on 
\be
r^2= \frac{\rho^2 +\alpha}{h}= \frac{\rho^2-a^2}{\beta^2 s(s-2)} \bigg( u \mp (s-1) \bigg)^{-2(s-1)} \frac{u\pm 1}{u\mp 1}\,.
\ee
Then Eq. \eqref{dzanddx} and Eq. \eqref{w0m1z} imply 
\be
\frac{\rho-a}{\rho+a} =e^{2a z_0 \pm u} \bigg( \frac{u\mp 1}{u\pm 1}  \bigg) \,.
\ee
We define $P=e^{2az_0}$ for the ``$+$'' branch and $P=-e^{2az_0}$  for the ``-" branch, and then write down $\rho$ as a function of $u$ in a unified expression
\be
\rho = a \frac{u+1 +P e^{u} (u-1) }{u+1 -P e^{u} (u-1) } \,.\label{w0m1x}
\ee 
Therefore, we have
\be
r^2=  \frac{4 a^2 P}{\beta^2 s(s-2)} \frac{ e^{2u}(u+s-1)^{-2(s-1)} (u+1)^2 }{\big( u+1-Pe^{2u}(u-1) \big)^2} \,. \label{w0m1S}
\ee
The solutions described by Eqs.~\eqref{hbtbyu}, Eq.~\eqref{w0m1x} and Eq.~\eqref{w0m1S} are summarized as Eq.~\eqref{BHfam}, which includes situations of black holes.

\subsection*{(II) The case of $ w_0 = - \frac{(s-1)^2}{s(s-2)}$}
We then study the following case
\be
w_0 = - \frac{(s-1)^2}{s(s-2)} \,.
\ee
It makes Eq.~\eqref{solforzofw} and Eq.~\eqref{solforpsiofw} become
\be
\frac{\der z}{\der w}=\frac{1}{2} \sqrt{-\frac{s(s-2)}{\alpha(1+w)} } \,,\quad 
\frac{\der \psi}{\der w}=\frac{s-1}{2w} \frac{1}{\sqrt{1+w} }  \,,
\ee
which can be integrated as
\begin{align}
	z=&~ z_0 \pm  \sqrt{-\frac{s(s-2)}{\alpha} } 	\sqrt{1+w} \,, \label{cancelcasez} \\
	\psi=&~ \psi_0 \pm \frac{s-1}{2}\log\bigg| \frac{ \sqrt{1+w}-1 }{ \sqrt{1+w}+1 } \bigg| \label{cancelcasepsi} \,.
\end{align}
Since $w= \big(\sqrt{1+w}-1 \big) \big(\sqrt{1+w}-1 \big)$,  Eq.~\eqref{cancelcasepsi} implies that
\begin{align}
	h=&~ \beta^2 \big(\sqrt{1+w}\pm 1 \big)^{2(s-1)}  	\,, \nonumber \\
	b_t =&~ \beta \sqrt{ \frac{2}{\kappa} } \big(\sqrt{1+w} \big) \big(\sqrt{1+w}\pm 1  \big)^{s-1} 	\,.
\end{align}
On the other hand, the factor $\sqrt{-s(s-2)/\alpha}$ in Eq.~\eqref{cancelcasez} demands $\alpha>0$ for the case of $0<s<2$, and
$\alpha<0$ for $s<0$ or $s>2$. It would be beneficial to distinguish these two situations.

If $0<s<2$, one can take $\alpha=a^2$ such that integrating out Eq. \eqref{dzanddx} leads to  
\be
z-z_0=  a^{-1}\bigg( \arctan\bigg( \frac{\rho}{a} \bigg)+ C\bigg) \,.
\ee
Then Eq.~\eqref{cancelcasez} implies
\be
\sqrt{1+w} = \frac{a(z-z_0)}{\sqrt{s(2-s)}} = \frac{1}{\sqrt{s(2-s)}}\arctan\bigg( \frac{\rho}{a}\bigg) + \frac{C}{\sqrt{s(2-s)}} \,,
\ee
in which we can always use $\rho$ or $a$ to absorb the sign. 
One can further choose $\beta=\pm [s(2-s)]^{(s-1)/2} $ or equivalently rescale $t$, $\rho$ and $a$ with suitable factors, and define $c=C\pm\sqrt{s(2-s)} $, and the corresponding $\{h(\rho), r^2(\rho), b_t(\rho)\}$ are
\begin{align}
	h(\rho) &=  \bigg( \arctan\bigg( \frac{\rho}{a} \bigg) +c \bigg)^{2(s-1)}  \,,\nonumber\\ 
	r^2(\rho) &=  \big(\rho^2+a^2\big)\bigg( \arctan\bigg( \frac{\rho}{a} \bigg) +c \bigg)^{-2(s-1)}  \,, \nonumber\\
	b_t(\rho) &= \sqrt{ \frac{2}{\kappa} }  \bigg( 1 \pm\frac{\arctan(\rho/a) +c  }{\sqrt{s(2-s)}}\bigg)
	\bigg(\arctan\bigg( \frac{\rho}{a} \bigg) +c \bigg)^{s-1} \,.\label{cancelcaseSolmal}
\end{align}
When $s=1$, the metric from Eqs.~\eqref{cancelcaseSolmal} reduces to the symmetric Ellis-Bronnikov (EB) wormhole \cite{Ellis:1973yv, Bronnikov:1973fh}.

If $s<0$ or $s>2$, one should take $\alpha=-a^2$. Then
Eq. \eqref{dzanddx} and Eq. \eqref{cancelcasez} imply
\be
a(z-z_0)= \bigg( \frac{1}{2}\log\bigg|\frac{\rho-a}{\rho+a}\bigg| +C\bigg) = \sqrt{s(s-2)}(\sqrt{1+w}) \,.
\ee
Following the same treatment, choosing $\beta=\pm [s(s-2)]^{(s-1)/2} $, one can take $\{h(\rho),r^2(\rho),b_t(\rho)\}$ as
\begin{align}
	h(\rho) &=   \bigg(\frac{1}{2}\log\bigg|\frac{\rho-a}{\rho+a}\bigg| +c \bigg)^{2(s-1)}  \,, \quad 
	r^2(\rho) = \big(x^2-a^2\big)\bigg(\frac{1}{2}\log\bigg|\frac{\rho-a}{\rho+a}\bigg| +c \bigg)^{-2(s-1)}   \,, \nonumber\\
	b_t(\rho) &= \sqrt{ \frac{2}{\kappa} } \bigg(1\pm  \frac{1}{2\sqrt{s(s-2)}}\bigg( \frac{1}{2}\log\bigg|\frac{\rho-a}{\rho+a}\bigg| +c  \bigg) \bigg)\; \bigg(\frac{1}{2}\log\bigg|\frac{\rho-a}{\rho+a}\bigg| +c \bigg)^{s-1}\,, \label{cancelcaseSolpal}
\end{align}
in which $c=C\pm \sqrt{s(s-2)}$.
\subsection*{(III)   Solutions for the case of $\xi=2\kappa$ }
It would be interesting to compare the case of $\xi=2\kappa$ (i.e., $s=2$) with the decoupling case $\xi=0$ (i.e., $s=0$). They share the same $\psi(w)$ and $z(w)$ (hence the same $\rho(w)$), but different $h(w)$ and $b_t(w)$.
It seems that constructing the coordinate transformation with $r$ and $w$ for the RN black hole naturally yields the exact solution for the case of $\xi=2\kappa$. However, it is not easy to seek the appropriate expression for such a coordinate transformation. Instead, the common $\{\psi(w),z(w)\}$ for $s=0$ and $s=2$ suggests choosing an alternative ansatz.

First, let us review the RN spacetime:
\be
\der s^2 = - \bigg( 1- \frac{2M}{r^2} + \frac{Q^2}{r^2} \bigg) \der t^2 
+  \bigg( 1- \frac{2M}{r^2} + \frac{Q^2}{r^2} \bigg)^{-1} \der r^2 +r^2\der\Omega^2  \,.
\ee
Simply comparing it with Eq.~\eqref{keyconstr}, we have $r^2h= r^2-2Mr+Q^2=(r-M)^2+Q^2-M^2 $.
To match our ansatz Eq.~\eqref{keyconstr}, we introduce
\be
\rho=r-M \, , \quad \alpha=Q^2-M^2 \,.
\ee
Hence the RN metric becomes
\be
ds^2 = - \frac{\rho^2+\alpha }{(\rho+M)^2} dt^2 
+   \frac{(\rho+M)^2}{\rho^2+\alpha } d\rho^2 +(\rho+M)^2d\Omega^2  \,.
\ee
It hints at the following new ansatz:
\begin{align}
	h(\rho) &= w^2(\rho) \hrn(\rho)=w^2(\rho)\frac{\rho^2+\alpha }{(\rho+\mrn)^2} \,, \nonumber\\
	r^2(\rho) &= \frac{r^2_{\text{RN}}(\rho)}{w(\rho)^2} = \frac{(\rho+\mrn)^2}{w(\rho)^2} \,,  \nonumber\\
	b_t(\rho) &= \pm\sqrt{\frac{2}{\kappa}}\, \frac{w(\rho)}{\rho+\mrn} \bigg( \big( 1+w(\rho)\big) \big( \rho^2+\alpha \big) \bigg)^{1/2}  \,,
\end{align}
where we introduce $\mrn$ to avoid misidentifying the ADM mass $M$ for the case of $\xi=2\kappa$ with the mass of the RN black hole.
Then the $t$-$t$ component of Eq. \eqref{eomgra} and the $t$ component of Eq. \eqref{eomvec} imply
\begin{align}
	0= 2 (\mrn^2 + \alpha) w + 2 (\rho+\mrn) (\rho^2 + 2 \mrn \rho  - \alpha) \frac{\der w}{\der \rho} + (\rho+\mrn)^2 (\rho^2 + \alpha) \frac{\der^2w}{\der \rho^2}
	\,,
\end{align}
and we obtain
\be
w(\rho)= \frac{ \rho+\mrn  }{\rho^2 +\alpha} ( C+ \beta \rho ) \,.
\ee
Therefore, we have
\begin{align}
	&\der s^2 = - \frac{(C+\beta \rho)^2}{\rho^2+\alpha } \der t^2 
	+   \frac{\rho^2+\alpha }{(C+\beta \rho)^2} \der\rho^2 +\bigg(\frac{{\rho^2+\alpha }}{C+\beta \rho}\bigg)^2 \der\Omega^2  \,,\nonumber \\
	&b_t(\rho)\der t =\pm\sqrt{\frac{2}{\kappa}}\, \frac{C+\beta \rho}{\rho^2+\alpha} \sqrt{\rho^2+\alpha+ (\rho+\mrn)(C+\beta \rho) } ~\der t\,.
\end{align}
However, it does not completely solve the EoMs.
Substituting the above results into Eq.~\eqref{GravityShsolved} and Eq.~\eqref{VecShsolved}, we obtain the following constraint on the parameters $\alpha$, $\beta$, $C$ and $\mrn$:
\begin{align}
C^2 - 2 C (2 + \beta) \mrn + \beta^2 \mrn^2 = 4 (1+\beta)\alpha \,.\label{xi2kappaConstrain}
\end{align}

To have solutions describing a spacetime with an asymptotically flat region, we should require $\beta\neq 0$.
Obviously, there is a special case satisfying Eq.~\eqref{xi2kappaConstrain}:
\begin{align}
  \beta = -1 \,,\quad
 \mrn=C\,,
\end{align}
where the corresponding
results are
\begin{align}
	&\der s^2 = - \frac{(\rho-C)^2}{\rho^2+\alpha } \der t^2 
	+   \frac{\rho^2+\alpha }{( \rho-C)^2} \der\rho^2 +\bigg(\frac{{\rho^2+\alpha }}{ \rho-C}\bigg)^2 \der\Omega^2  \,,\nonumber \\
	&b_t(\rho)\der t =\mp\sqrt{\frac{2}{\kappa}}\, \frac{ \rho -C}{\rho^2+\alpha} \sqrt{\alpha+ C^2 } ~\der t\,. \label{xi2kappabetam1}
\end{align}

For the case of $\beta\neq -1$,
one can introduce $\alpha =  \beta^2 \bar{\alpha} $ and rescale $t=\bar{t}/\beta$, $\rho=\beta \rhobar$, $\mrn=\beta^2\mu $ and $C=\beta^2\,\nu$, such that Eq.~\eqref{xi2kappaConstrain} becomes
\begin{align}
\nu^2 - 2  (2 + \beta) \nu\mu + \beta^2 \mu^2 = 4 \bar{\alpha} \frac{1+\beta}{\beta^2} \,.\label{xi2kappaConstrain}
\end{align}
and the solutions become
\begin{align}
	&\der s^2 = - \frac{(\rhobar + \nu )^2}{\rhobar^2+\bar{\alpha} } \der\bar{t}^2 
	+ \frac{\rhobar^2+\bar{\alpha}}{(\rhobar + \nu )^2} \der\rhobar^2 +\bigg(\frac{{\rhobar^2+\bar{\alpha} }}{\rhobar+\nu}\bigg)^2 \der\Omega^2  \,,\nonumber \\
	&b_t \der t= \pm\sqrt{\frac{2}{\kappa}}~\frac{\rhobar + \nu}{\rhobar^2 + \bar{\alpha} } \bigg( \sqrt{(1+\beta) }~\rhobar+\frac{  \beta(\nu+\beta\mu) }{2\sqrt{(1+\beta) }}\bigg)  ~\der\bar{t}\,.
\end{align}
Hence these two cases contain a common formula for the line-element, though the bumblebee vector field configurations are different.

Moreover, the stealth Schwarzschild solution can appear in the situation of $\alpha<0$, if $\nu$ takes an appropriate value. For simplicity, we introduce $\alpha=-a^2$ and consider $\nu=a$. Hence, the common factor $(\rhobar+a)$ in $(\rhobar+a)^2$ and $\rhobar-a^2$ cancels. The metric becomes
\begin{align}
	&\der s^2 = - \frac{\rhobar + a }{\rhobar - a } \der\bar{t}^2 
	+ \frac{\rhobar - a }{\rhobar + a} ~\der\rhobar^2 +\big(\rhobar -a\big)^2 d\Omega^2  \,.
\end{align}
Then, shifting $r=\rhobar-a$, one can recover the usual expression of the Schwarzschild metric
\begin{align}
	&ds^2 = - \frac{r - 2a }{r } d\bar{t}^2 
	+ \frac{r }{r -  2a} dr^2 +r^2 d\Omega^2  \,.
\end{align}
The case of $\nu=-a$ also leads to the stealth Schwarzschild solution.
However, choosing $C=\pm a$ for Eqs.~\eqref{xi2kappabetam1} yields the Schwarzschild solution, rather than the stealth Schwarzschild solution, because the bumblebee vector field vanishes due to $\alpha+C^2=0$. 
\end{widetext}

\bibliography{bum_refs} 
\end{document}